\documentclass{aa}
\usepackage{graphicx}
\usepackage{natbib}
\usepackage{aalongtable}
\usepackage{rotating}
\usepackage{txfonts}

\bibpunct{(}{)}{;}{a}{}{,}
\def\aap{A\&A}
\def\aaps{A\&AS}
\def\aj{AJ}
\def\apj{ApJ}
\def\apjs{ApJS}
\def\apss{Ap\&SS}
\def\apjl{ApJ}

\def\mnras{MNRAS}
\def\nat{Nature}
\def\pasp{PASP}
\def\pasj{PASJ}

\def\aapr{A\&ARev}
\def\memras{Mem.R.Astron.Soc.}
\def\pasa{PASA}
\def\jrasc{JRASC}
\def\iaucirc{IAU Circ.}
\def\na{New Astron.}
\def\rmxaa{Rev.Mex.A.A.}

\begin{document}

\title{Catalogue of particle-accelerating colliding-wind binaries} 
%subtitle{}

\author{M. De Becker \and F. Raucq}

\offprints{M. De Becker}

\institute{Department of Astrophysics, Geophysics and Oceanography, University of Li\`ege, 17, All\'ee du 6 Ao\^ut, B5c, B-4000 Sart Tilman, Belgium \\
}

\date{Received ; accepted }

\abstract
{Massive systems made of two or more stars are known to be the site for interesting physical processes -- including at least in some cases -- particle acceleration. Over the past decade, this topic motivated a particular effort to unveil the properties of these systems and characterize the circumstances responsible for the acceleration of particles and the potential role of pre-supernova massive stars in the production of high energy particles in our Galaxy. Although previous studies on this topic were mostly devoted to processes in general, or to a few individual objects in particular, a unified target-oriented census of particle-accelerating colliding-wind binaries (hereafter PACWBs) does not exist yet. This paper aims at making a general and unified census of these systems, emphasizing their main properties. A general discussion includes energetic considerations along with wind properties in relation with non-thermal emission processes that are likely at work in colliding-wind binaries. Finally, some guidelines for future observational and theoretical studies are drawn.}


\authorrunning{De Becker \& Raucq}
\titlerunning{Catalogue of PACWBs}

\maketitle

\section{Introduction}
\label{intro}

The study of massive stars across the electromagnetic spectrum revealed the existence of many physical phenomena lying at the core of modern stellar astrophysics. Despite their relative low number, massive stars play a significant role in shaping their environment and feeding the interstellar medium with material, kinetic energy, and radiation. In addition, the identification of synchrotron radio emission from a few massive stars about thirty years ago provided evidence that some of them were able to accelerate particles up to relativistic energies \citep{WhBe,ABC,Wh}. This discovery contributed to open a new window in the investigation of massive star's physics: Non-degenerate, early-type stars appeared to be very energetic sources that are able to accelerate particles up to relativistic energies, therefore potentially contributing to the production of cosmic-rays in our Galaxy.

Since then, many studies contributed to improve our view of the circumstances responsible for the efficiency of the particle acceleration process. Most of the related studies aimed at investigating the properties of the radio emission of massive stars to identify hints of non-thermal emission processes: radio spectral index, brightness temperature, and/or variable radio emission \citep[see][]{WhBe,ABC,DW,debeckerreview}. These studies led to a census of {\it non-thermal radio emitters}, which are massive stars whose radio emission presents a significant departure from the canonical thermal behaviour. It is however important to emphasize here that non-thermal radio emission should only be considered as an observational evidence, potentially biased by particular circumstances, such as free-free absorption, which prevents us to detect the synchrotron radio radiation. Consequently, we cannot a priori reject a scenario where particles are indeed accelerated up to relativistic energies, even in objects that are not tagged as non-thermal radio emitters. In addition, the presence of relativistic particles admits non-thermal emission processes to be at work in the high energy domain as well \citep{PD140art,Pit2,DSACWB,debeckerreview}, as recently detected in the case of two long-period massive binaries: $\eta$\,Car \citep{viottietacar,tavanietacar} and WR\,140 \citep{sugawara140}. Both non-thermal emission processes in the radio and in the high energy domains therefore appear as two independent and indirect evidences for the existence of a population of relativistic particles. The intrinsic phenomenon worth investigating is the particle acceleration process at work in massive stars. Finally, one should also emphasize that most objects responsible for non-thermal emission processes appear to be binaries, or even higher multiplicity systems \citep[see][and Sect.\,\ref{multisection}]{debeckerreview,benagliareview}. Consequently, we consider in the context of this paper that the category of non-thermal radio emitters is definitely superseded by that of {\it particle-accelerating colliding-wind binaries} (PACWBs), a concept that is more physically justified than that of non-thermal radio emitter. The recent convergence of theoretical expectations (prediction of non-thermal high-energy emission) and observations (such as detections in X-rays and $\gamma$-rays), and significant progress in the observational exploration of this category of objects in the past years, justifies an updated census related to these objects, which is the motivation of the present catalogue in which all relevant objects are presented following an unified approach, whatever their spectral classification.

As a first step, the catalogue is presented in Sect.\,\ref{catasection}, and the important issue of the multiplicity is addressed with more details in Sect.\,\ref{multisection}. A census of important observational results in the radio and high energy domains is presented in Sects.\,\ref{radiosection} and \ref{highenergysection}, respectively. These results are discussed in Sect.\,\ref{discussionsection}, where some guidelines for future investigations related to particle acceleration in colliding-wind binaries are also formulated. Finally, Sect.\,\ref{conclusionssection} enunciates some conclusions. 

\section{The catalogue}\label{catasection}

In the context of particle acceleration by stellar sources, especially by massive stars, a few words of clarification are needed. First, it is emphasized that this paper focuses on massive non-degenerate stars in binary or higher multiplicity systems, leading to a wind-wind interaction region characerized by two shocks likely responsible for the acceleration of particles through the diffusive shock acceleration (DSA) mechanism \citep{PD140art,DSACWB,debeckerreview}. As a result, this study does not concern massive X-ray and/or $\gamma$-ray binaries or microquasars that include a compact object \citep{fenderhk,dmitry2005,valenti2007,valenti2011} and is neither related to supernova remnants, known to be efficient particle accelerators \citep{romerosnr,reynoldssnr,vinkSNRutrecht}. Moreover, the class of objects debated does not include rapidly-moving runaway massive stars whose stellar wind interacts with the interstellar medium, producing strong shocks likely to accelerate particles \citep{benagliarunaway,romeroliac,peri2011,ntbowshock}.

The current census of massive binaries (or at least suspected binaries) displaying observational evidence for the existence of a population of relativistic electrons is presented in Table\,\ref{cata}. The table, referred to as {\it List A} (43 objects), includes confirmed and strongly suspected binary/multiple systems, along with 6 objects whose multiplicity has never been investigated so far (see Sect.\,\ref{multisection} for details on the multiplicity of the objects considered). In each case, an usual and an alternative identifier are provided, along with the equatorial coordinates (J2000) retrieved from the SIMBAD Astronomical Database\footnote{\tt http://simbad.u-strasbg.fr/simbad/}. The distance to each system is also given with the related bibliographic references. One should note that the current most accepted value for the distance is given for most of the time. However, the issue of the value for the distance, in some cases, is still uncertain, and a widely accepted value is provided. In these cases, the specified reference consists of a review of the literature making a census of the various distance determinations so far. Finally, the given value for the distance, in many cases, is that of the related open cluster or OB association, provided the membership is not critically disputed.

In contrast to a previous census of massive stellar non-thermal emitters \citep{DW,debeckerreview,benagliareview}, it is important to note that WR-type and OB-type objects are not separated. This decision is motivated by the idea that whichever object or the evolutionary stage that is considered, one is dealing with the same physics. The only difference between systems included in this catalogue arise from different locations in the stellar, wind, and orbital parameter space. Second, the strict distinction between WR and OB stars does not appear to be relevant, considering the uncertainties on the spectral classification of some objects (e.g. the primary stars in WR\,21a and Cyg\,OB2\,\#5, or the most enigmatic colliding-wind binary $\eta$\,Car). 

So far, 43 objects belong to the class of non-degenerate early-type stars that give rise to an efficient particle acceleration process in a wind-wind interaction region. Most of the members of this catalogue consist of non-thermal radio emitters, whose membership is mainly attributed on the basis of the criteria detailed in Sect.\,\ref{radiosection}. However, complementary criteria, such as the detection of non-thermal high energy emission, are also considered in this work, following recent discoveries in X-rays and $\gamma$-rays, as discussed in Sect.\,\ref{highenergysection}. In addition, membership in the class of PACWBs is assumed for some objects, despite the lack of firm evidence of the presence of a companion. As a rule, a non-thermal emitter is considered as a member in the catalogue, provided that there is no obvious reason that has been found in the literature to reject it. As a result, a few objects previously classified as (potential) non-thermal radio emitters are not included in the present catalogue for reasons explained in Sect.\,\ref{radiosection} ({\it List B}). Some of these objects may be included later on, depending on significant clarification of their status. Individual comments about the members of {\it Lists\, A} and {\it B} are provided in Appendix\,\ref{individualA} and Appendix\,\ref{individualB}, respectively.

%Multiple citations to the same reference:
%\newcounter{test} \setcounter{test}{\value{refmulti}} at first citation
%\thetest at next citations 

\section{Multiplicity}\label{multisection}

As already pointed out by previous studies \citep{DW,BRwr,EU,sven2006,debeckerreview,benagliareview}, multiplicity seems to play a determining role in the particle acceleration process at work in massive stars. It is therefore crucial to address first the issue of the multiplicity. The current information on the multiplicity of objects included in this catalogue is given in Table\,\ref{multi}, along with the spectral classifications of the components in the binary -- or higher multiplicity -- systems. It is striking to see that 37 out of the 43 objects (more than 85\,$\%$) are confirmed or at least suspected binaries, among which only 5 are not firmly confirmed binaries. Objects with undetermined multiplicity status (6 out of the 43) are still lacking dedicated observations in spectroscopy or high angular resolution imaging, which aim at investigating the potential presence of companions. The lack of identified companion for these objects should not therefore be considered as evidence of a single star status. As a result, the binary fraction of at least 85\,$\%$ should be considered as a lower boundary of the actual value. On the other hand, it should be noted that the binary fraction among massive stars, whatever their classification, should not be larger than a conservative limit of 50\,$\%$ \citep{sanaevans}. As a consequence, one can comfortably claim that the high binary fraction among PACWBs is specific to this subsample of massive stars that is not attributable to the general population of massive stars. This observational result lends significant support to the {\it standard scenario} for PACWBs, where binarity is a strong requirement to detect non-thermal emission (whatever the spectral domain, radio or high energies).

The binary/multiple nature of a high fraction of the objects in Table\,\ref{multi} has been revealed by spectroscopic studies. For a few systems, the presence of a companion has been provided by high angular resolution observations, such as in the case of HD\,93250. The investigation of the multiplicity of massive stars is not so straightforward, and a good complementarity between spectroscopy and imaging is required to get rid of the numerous observational biases that affect our capability to attribute adequate multiplicity status to massive stars \citep{sanaevans}. In addition, the presence of the companion is strongly suspected in the case of some systems but not yet confirmed. Indirect evidence of binarity considered in Table\,\ref{multi} are variable and potentially periodic radio emission or, to some extent, hard and variable X-ray emission. Considerations related to radio and X-ray emission will be discussed in Sects.\,\ref{radiosection} and \ref{highenergysection}.

Although most of the objects in the catalogue are identified as non-single stars, it is highly relevant to make a census of the systems for which orbital elements have been confidently derived so far, as illustrated in Table\,\ref{orbsol}. The geometrical aspects of the system (orbit size, stellar separation, eccentricity, etc) indeed play a significant role in the particle acceleration mechanism and in the related non-thermal radiation processes. The lack of orbital solution for many systems in the catalogue calls upon dedicated studies using spectroscopic and interferometric techniques.

\begin{table}
\caption{List of systems with existing orbital solution.\label{orbsol}}
\begin{center}
\begin{tabular}{l l l}
\hline
\# & usual ID & Reference for orbital elements \\
\hline
\vspace*{-0.2cm}\\
1 & HD\,15558 & \citet{ic1805_2} \\
4 & 15\,Mon & \citet{gies15mon} \\
6 & WR\,11 & \citet{gam2velorbsol1,gam2velorbsol} \\
9 & WR\,21a & \citet{wr21aniemela} \\
16 & HD\,150136 & \citet{sana150136} \\
27 & 9\,Sgr & \cite{rauw9sgr2} \\
35 & WR\,133 & \citet{UH1994wr133} \\
36 & WR\,137 & \citet{wr137orbit} \\
37 & WR\,140 & \citet{fahed140mnras} \\
39 & Cyg\,OB2\,\#9 & \citet{Blommecyg9} \\
40 & Cyg\,OB2\,\#8A & \citet{Let8a} \\
\vspace*{-0.2cm}\\
\hline
\end{tabular}
\end{center}
\end{table}

\section{Radio investigations}\label{radiosection}

As already mentioned, most PACWBs have been identified through synchrotron radiation in the radio domain (produced in addition to their free-free radiation). The thermal radio emission from massive stars is produced in their stellar winds. The radio flux density (S$_\nu$, generally expressed in mJy) presents a power law dependence on the frequency ($\nu$) of the type $\mathrm{S}_\nu \propto \nu^\alpha$ with a spectral index $\alpha$ close to 0.6 for a free-free radio emission from a spherically symmetric and homogeneous, single star wind \citep{PF,WB}. In the case of a binary system with interacting winds, a small thermal radio contribution can be produced by the shocked plasma in the interaction region \citep{Pit2}. In the standard model of the particle acceleration in massive binaries, the non-thermal radio emission (i.e., synchrotron radiation) is produced in the presence of a population of relativistic electrons accelerated by shocks in the wind-wind interaction region.

\newcounter{id}
\setcounter{id}{0}
\begin{table*}
\caption{Candidates for non-thermal emission not included in the catalogue, as referred to as {\it List B}.\label{cand}}
\begin{center}
\begin{tabular}{l l l l l l l}
\hline
\# & usual ID & other ID & $\alpha$[J2000] & $\delta$[J2000] & Sp. type(s) & Remark \\
\hline
\vspace*{-0.2cm}\\
\stepcounter{id}\theid & CC\,Cas & \object{HD~19820} & 03 14 05.34 & +59 33 48.48 & O8.5III + B0V & Possible eclipse-like variation\\
\stepcounter{id}\theid & $\xi$\,Per & \object{HD~24912} & 03 58 57.90 & +35 47 27.71 & O7.5III (or I) + ? & Runaway star\\
\stepcounter{id}\theid & $\alpha$\,Cam & \object{HD~30614} & 04 54 03.01  & +66 20 33.64 & O9.5I & Runaway candidate \\
\stepcounter{id}\theid & $\delta$\,Ori\,C & \object{HD~36485} & 05 32 00.41 & --00 17 04.46 & B2V & Magnetic Bp star \\
\stepcounter{id}\theid & $\theta^1$\,Ori\,A & \object{HD~37020} & 05 35 15.85 & --05 23 14.34 & B0.5V + TT + ? & Presence of a T\,Tauri component \\
\stepcounter{id}\theid & \object{HD~37017} & HR\,1890 & 05 35 21.87 & --04 29 39.02 & B1.5IV + ? & Magnetic Bp star \\
\stepcounter{id}\theid & $\sigma$\,Ori\,E & \object{HD~37479} & 05 38 47.19 & --02 35 40.54 & B2V & Magnetic Bp star \\
\stepcounter{id}\theid & \object{WR~22} & HD\,92740 & 10 41 17.52 & --59 40 36.90 & WN7 + O9 & $\alpha$ not accurate \\
\stepcounter{id}\theid & \object{WR~79} & HD\,152270 & 16 54 19.70 & --41 49 11.54 & WC7 + O5-8 & Possible source confusion \\
\stepcounter{id}\theid & \object{WR~86} & HD\,156327 & 17 18 23.06 & --34 24 30.63 & WC + OB & $\alpha$ not accurate \\
\stepcounter{id}\theid & W43\,\#1 & -- & 18 47 36.69 & --01 56 33.06 & WR? + ? & Possible phenomena combination \\
\stepcounter{id}\theid & Cyg\,OB2\,\#11 & \object{BD~+41 3807} & 20 34 08.52 & +41 36 59.36 & O5I(f) & NT status uncertain\\ 
\stepcounter{id}\theid & \object{WR~156} & HIP\,113569 & 23 00 10.12 & +60 55 38.42 & WN8h + OB & NT status uncertain \\
\vspace*{-0.2cm}\\
\hline
\end{tabular}
\end{center}
\end{table*}
\normalsize

Radio observations can provide a wealth of information related to the nature of the emission processes at work in a massive binary:
\begin{enumerate}
\item[-] {\it The spectral index.} Any significant deviation with respect to the canonical thermal index is generally considered as a valuable tracer of non-thermal emission. Typically, non-thermal radio emission is associated with spectral indices significantly lower than 0.6, or even negative. It should be pointed out that the small thermal contribution from the wind-wind interaction itself is likely to cause the spectral index to be lower than 0.6, even in the absence of synchrotron radiation \citep{Pit2}. For this reason, $\alpha$ values between about +0.3 and +0.6 cannot firmly be associated with synchrotron radiation. The identification of particle accelerators on the basis of the spectral index value therefore requires flux density measurements of at least at two different frequencies. This criterion is unfortunately not met for several massive stars already observed in the radio domain \citep[see][for additional information]{benagliareview}.
\item[-] {\it The brightness temperature.} This quantity can be defined as the temperature that a black body would have to produce the same flux density as the radio source considered\footnote{Brightness temperature is defined in the context of the Rayleigh-Jeans limit (low frequencies) of Planck's function, for an optically thick source.}. Non-thermal radio emitters are expected to be characterized by brightness temperatures (T$_\mathrm{B} \sim 10^6 - 10^7$\,K), which are significantly higher than purely thermal emitters (T$_{\mathrm B} \sim 10^4$\,K). However, it should be emphasized that the strong variability that is expected for many PACWBs may cause the flux density (and therefore the corresponding T$_\mathrm{B}$) to reach low levels in a significant fraction of the orbit. As a result, massive binary systems that are able to accelerate particles may not be identified through this criterion, if the orbit is observationally poorly sampled.
\item[-] {\it Radio variability.} As the thermal radio emission is dominated by individual winds of the system that should be constant (at least on time scales significantly shorter than evolution time scale), the thermal emission is not expected to be intrinsically variable as a function of time. However, the binary/multiple nature of PACWBs introduces an obvious variability time scale, namely the orbital period. Considering the typical orbital periods (greater than a few weeks, see Sect.\,\ref{multisection}), even partial eclipses leading to periodic changes in the free-free absorption by the stellar winds are unlikely. However, the synchrotron radiation produced in the wind-wind interaction region is expected to be intrinsically variable if the orbit is eccentric because of the periodic change in the stellar separation. In addition, the synchrotron emission region is also very sensitive to orientation effects since the line of sight crosses varying absorbing columns as a function of the orbital phase. For instance, the existence of a companion star has been suggested, or confirmed, on the basis of such a variability in the cases of HD\,168112 \citep{DeB168112,Blo168112}, HD\,167971 \citep{Blo167971} and Cyg\,OB2\,\#5 \citep{kennedycyg5}. However, a significantly variable radio emission -- even as a function of the orbital phase -- is not enough to identify the non-thermal nature of the radio photons. A striking example is that of the enigmatic colliding-wind binary $\eta$\,Car, whose phase-dependent radio light curve is notably explained by varying ionization conditions in the wind primary \citep{etacarradio}. Even though we know it is an efficient particle accelerator (see Sect.\,\ref{highenergysection}), $\eta$\,Car is indeed the only known PACWB that is not a non-thermal radio emitter. Finally, one should note that the change in the relative importance of thermal and non-thermal contributions, combined with the modulation of the radio emission by free-free absorption as a function of the orbital phase, are expected to lead to changes in the measured radio spectral index as well. In eccentric massive binaries, the synchrotron emitting region is indeed expected to be embedded at various depths in the radio opaque stellar winds, depending on the orbital phase. As the free-free absorption is frequency dependent, the emerging flux density distribution (i.e., the spectral index) will depend on the orbital phase. The potentially variable nature of the radio emission emphasizes the interest to explore the temporal dimension in observational studies: Multiple observations are able to carry information that is not accessible through a snapshot observation.
\item[-] {\it High angular resolution imaging.} In a few cases, very long baseline radio interferometer arrays are allowed to resolve individual stellar winds (producing the thermal radiation) and the non-thermal emission region: HD\,93129A \citep{benagliahd93129avlbi}, Cyg\,OB2\,\#5 \citep{contr,ortizcyg5}, Cyg\,OB2\,\#9 \citep{vlbi2006}, WR\,146 \citep{oconnorwr146art}, WR\,147 \citep{williamswr147} and WR\,140 \citep{Doug140}. These results provide significant evidence that the synchrotron emission region is spatially coincident with the wind-wind interaction region, lending further support to the standard scenario for PACWBs.
\end{enumerate} 

The first three criteria constitute valuable hints for non-thermal radio emission. Our selection of catalogue members is mostly motivated by the fulfilment of at least one, or ideally two, of these criteria. The main information on the catalogue members resulting from radio studies is summarized in Table\,\ref{radio}. In this table, the third column specifies whether the spectral index ($\alpha$) has been measured to be negative or between 0.0 and 0.3 at least once. The same column also specifies whether the flux density at a given frequency has been reported to be variable. The relevant references are specified in the fourth column. Depending on the available information about the reported non-thermal emission, one should realize that all these objects are not characterized by the same level of confirmation about their non-thermal emitter status. A few objects present indeed hint for synchrotron radio emission on the basis of a low (but still positive) spectral index, or a suggested variable emission. For this reason, an arbitrary quality flag (Q in the last column of Table\,\ref{radio}) is attributed to each object: I and II stand, respectively, for a {\it certain} and {\it likely} non-thermal emitter status. We should also mention that other candidates for non-thermal emission processes in the radio domain are suggested in the literature. We refrained, however, from including them in the catalogue ({\it Lists A}) because of an unclear non-thermal emitter status or because the object itself presents some peculiarities, suggesting it does not adequately fit the colliding-wind binary scenario. These candidates are enumerated in Table\,\ref{cand}. Additional information about the reason for not including these sources in the catalogue are provided in Appendix\,\ref{individualB}. Among these objects, we emphasize the interesting cases of WR\,22 and WR\,86. The former is a known sepctroscopic binary with a period of about 80\,d \citep{binary1wr22,binary2wr22}, while the latter is a visual binary whose period (if gravitationally bound) should be very long \citep{binarywr86}. The two objects present a spectral index between 0.0 and 0.6, but the error bars on the spectral index do not allow to firmly establish the non-thermal nature of at least a fraction of the radio emission \citep{DW}. More accurate flux density measurements at different radio frequencies are needed to clarify the status of these objects. It is important to note that members of {\it List\,B} may be moved to {\it List\,A} in the future, depending on the results of new and expected investigations aiming at clarifying their nature and properties.

\section{High-energy investigations}\label{highenergysection}

The existence of a population of relativistic electrons opens the possibility triggering non-thermal emission processes in the high energy domain as well \citep{EU,BRwr,PD140art,DSACWB,debeckerreview}. One may distinguish between leptonic and hadronic processes. The first category involves electrons, with the most efficient process being inverse Compton (IC) scattering in the presence of the strong UV/Visible radiation field produced by the massive stars. Leptonic processes are expected to be able to produce high energy photons up to a few GeV, depending on the system properties. The second category involves notably relativistic protons, whose interaction with un-accelerated material leads to pion production, followed by decay. This hadronic process is efficient at producing $\gamma$-rays.

Considering that massive binaries are known to be bright thermal X-ray emitters \citep[see e.g.][]{SBP,pitpar2010}, the X-ray spectrum is expected to be dominated by thermal emission in the complete soft band (i.e. below 10\,keV), which agrees with observational results obtained over the past decades. In that energy band, the X-ray emission is indeed dominated by emission lines on top of a free-free continuum, as produced by an optically thin plasma that is characterized by temperatures that do not go beyond several 10\,MK (for CWB with periods of at least a few weeks, where stars are able to accelerate their winds up to their terminal velocitites, see \citealt{SBP,PS,pitpar2010}). As a result, one could only expect non-thermal high energy emission to be revealed in the hard X-ray domain (above 10\,keV) or even in $\gamma$-rays, where thermal X-rays are not present. We refer to \citet{debeckerreview} for a detailed discussion of this issue. As a result, any firmly established detection of high energy photons above 10\,keV from a colliding-wind binary could be considered as strong evidence of the occurrence of an efficient particle acceleration process in the system: This constitutes a solid criteria to belong to the class of PACWBs. Although predicted for many years, this expectation was confirmed by observations only a few years ago in the case of two long period massive binaries:
\begin{enumerate}
\item[-] $\eta$\,Car is the only CWB that displays non-thermal emission in both hard X-rays and $\gamma$-rays without any synchrotron detection in the radio domain. 
\item[-] WR\,140 is the only CWB simultaneously presenting radio synchrotron radiation and non-thermal hard X-rays.
\end{enumerate}
These two observational results shed new light on the physical processes at work in colliding-wind binaries and triggers additional observational and theoretical efforts in relation with this class of objects.\\

An overview of the main observational results related to objects in {\it List A} in the high energy domain is given in Table\,\ref{highenergy}. Even though hard X-rays and $\gamma$-rays are relevant as far as non-thermal emission processes are concerned, it is interesting to make a census of soft X-ray detections. First, crucial information on the multiplicity can be obtained through thermal X-rays, since the presence of colliding winds is expected to provide a significant additional emission component with respect to single massive stars. Systems for which the multiplicity is not yet fully established may unveil their companion through a significant X-ray variability, an overluminosity with respect to expected single stars, or even a hard thermal X-ray spectrum, which accounts for higher palsma temperatures\footnote{Colliding-wind binaries present pre-shock velocities that can reach the terminal velocity value, which are typically 2000--3000\,km\,s$^{-1}$. However, intrinsic shocks in individual stellar winds resulting from the line-driving instability, are characterized by pre-shock velocities at several 100\,km\,s$^{-1}$ \citep{FeldX}, which therefore leads to softer thermal X-ray spectra than in CWBs.}. Second, the detailed modelling of the physics of colliding-winds, including particle acceleration physics, requires a good knowledge of the hydrodynamics of the wind-wind interaction phenomenon, that is intimately related to the thermal X-ray emission from these objects. Soft X-ray observations are therefore very important to understand the physics of PACWBs.\\

In Table\,\ref{highenergy}, most catalogue members appear to be soft X-ray emitters. Among systems known to be at least binaries, only WR\,39, WR\,98, WR\,98a, WR\,112, Cyg\,OB2-335, and WR\,146 have never been detected below 10\,keV. It is important to note that the lack of clear detection of soft X-rays from these systems is not correlated with the absence of a known companion. WR\,98, WR\,98a, Cyg\,OB2-335, and WR\,146 are confirmed binaries that cover a wide range of orbital periods, and WR\,112 is strongly believed to be a binary system, even though a firm clarification of its multiplicity status is still lacking. Among the few stars with undetermined multiplicity status, no soft X-ray detection has been reported for CD\,--47\,4551, WR\,90, and HD\,190603. In the case of Wolf-Rayet stars with strong stellar winds (typically, WC-stars), the thickness of the wind material is known to be able to significantly absorb soft X-rays, seriously affecting their probability to be detected. It should also be noted that many objects in the catalogue never benefitted from a dedicated X-ray observation with the most recent and sensitive facilities (i.e. XMM-Newton, Chandra, Suzaku), and the information on soft X-ray emission from these objects comes mainly from less sensitive satellites, such as EINSTEIN or ROSAT. These last observatories operated only below 2.5\,keV, which is the energy range that is the most affected by photoelectric absorption by stellar wind and interstellar material. Current observations of these objects are therefore strongly biased, and dedicated observations are needed to clarify their X-ray emitting status below 10\,keV.

At higher energy wavebands (above 10\,keV), only a few objects have been investigated using INTEGRAL and SUZAKU, leading to a couple of detections (WR\,140 and $\eta$\,Car). However, the sensitivity of current observatories is not good enough to detect most PACWBs in the hard X-ray domain, which agrees with the energy budget considerations developed in Sect.\,\ref{discussionsection}. The same is true at $\gamma$-ray energies. In the context of $\gamma$-ray emission related to massive stars, one could also mention the case of the Fermi source 2FGL\,J2030.7+4417 \citep{fermicata2} that is coincident with the massive runaway O + B binary \object{HD195592} \citep{DeBhd195592}. However, the most probable scenario explaining the $\gamma$-ray emission, provided it is associated with HD\,195592, is non-thermal emission from the bow shock produced by the interaction between the wind material and the interstellar medium \citep[for a general model for this scenario, see][]{ntbowshock}. The investigation of this scenario specifically applied to the case of HD\,195592 can be found in \citet{DRD2013}.

\section{Discussion}\label{discussionsection}

\subsection{Parameter space}\label{paramspace}

\begin{figure*}[ht]
\begin{center}
\includegraphics[width=160mm]{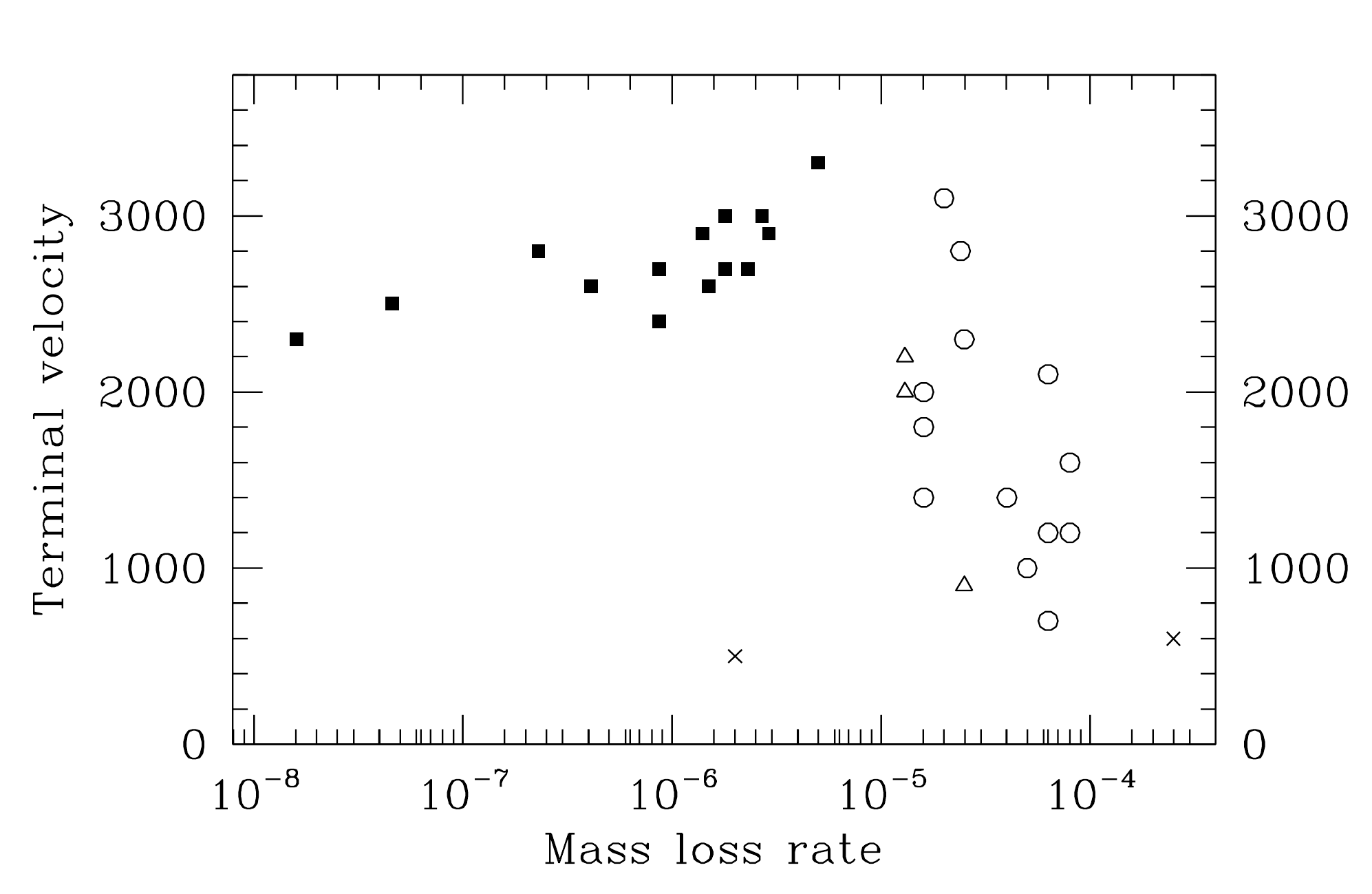}
\caption{Coverage of PACWBs in the ${\dot M}-V_\infty$ plane. Filled squares stand for O-type stars, open circles for WR stars, open triangles for Of/WN objects, and crosses for LBV-like objects. Mass loss rates are expressed in M$_\odot$\,yr$^{-1}$ and terminal velocities in km\,s$^{-1}$.\label{mdotvterm}}
\end{center}
\end{figure*}

From the current census of objects belonging to the category of PACWBs, one may have a look at the parameter space covered by these objects. In this context, the mass loss rate and the terminal velocity are especially important. The adopted mass loss rates and terminal velocities in each case are given in the fourth and fifth columns of Table\,\ref{param}. The quoted values are those characterizing the most powerful wind in the system\footnote{In the limited case of a perfectly symmetric system, the assumption to consider the contribution from one of the two objects leads only to a factor 2 for the error in the determination of the kinetic power based on the mass loss rates and terminal velocities of the dominant wind. This is reasonable considering the approximation level of the approach followed here. Unfortunately, the nature of the companion(s) is not well established for many objects, therefore preventing a similar analysis for all stars harboured in these systems to be performed.}. For O-type stars, we used the predicted (theoretical) mass loss rates and terminal velocities given by \citet{predmdot}, except for the terminal velocity of main-sequence O-stars: We considered V$_\infty$ = 2.6\,V$_\mathrm{esc}$, where V$_\mathrm{esc}$ is the escape velocity given by \citet{predmdot}, because their quoted terminal velocities seem unexpectedly high for this luminosity class. In the case of WN and WC type stars, we considered mean values for the mass loss rate and for the terminal velocities quoted by \citet{HK1998WN} and \citet{SHT2012WC}, respectively, for a given spectral type, except for objects specifically mentioned in these studies. In the cases of objects considered to be intermediate between Of and WN stars, an intermediate value between O supergiants and WN values was adopted. In the absence of specific and reliable quantities, our assumption is that the values adopted in this study are typical of the stars considered. In the specific cases of $\eta$\,Car, WR\,140, and HD\,190603, we selected the parameters given by \citet{PCetacar}, \citet{WilliamsLIAC} and \citet{clarkBHG}, respectively. The adopted mass loss rates and terminal velocities in each case are given in Table\,\ref{param}.  To explore the parameter space covered by the {\it list\,A} sample, we plotted in Fig.\,\ref{mdotvterm} the position of each object in the ${\dot M}-V_\infty$ plane. It is noticeable that objects with the lowest mass loss rates are not those characterized by the lowest terminal velocities, expectedly translating the need for a sufficiently high kinetic power to significantly feed non-thermal processes. Mass loss rates cover about four orders of magnitude, which agrees with the wide range of spectral classifications of PACWBs. The lower left part of the ${\dot M}-V_\infty$ diagram corresponds to the parameter space, which is more typical of B-type stars whose kinetic power is probably not high enough to give rise to detectable non-thermal emission processes. 

\begin{figure*}[ht]
\begin{center}
\includegraphics[width=160mm]{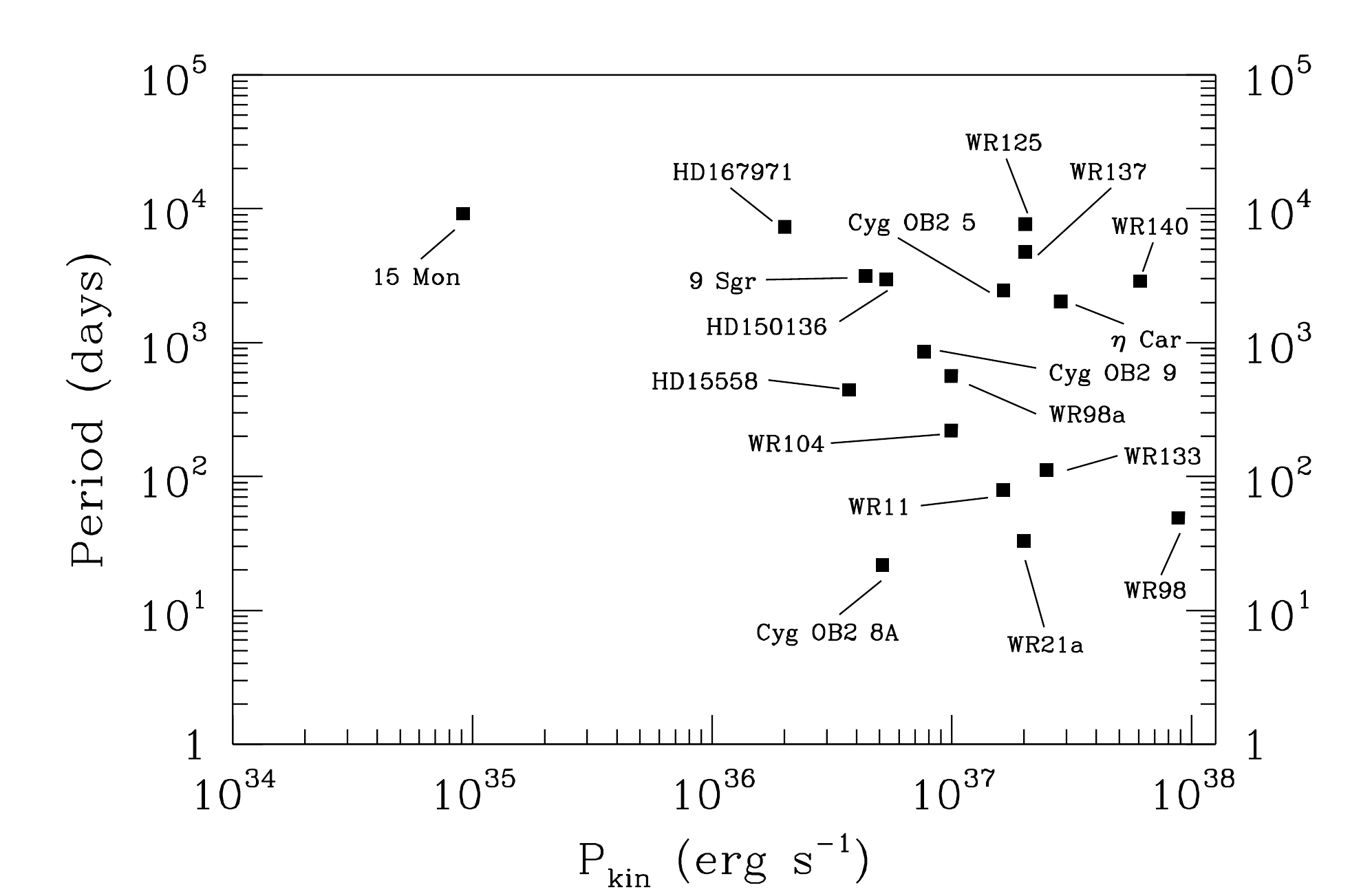}
\caption{Coverage of PACWBs in the Period$-P_{kin}$ plane for objects whose orbital period is determined. \label{period}}
\end{center}
\end{figure*}

These quantities are important as they define the kinetic power of a stellar wind, which is a pivotal quantity for energy budget considerations as developed in Sect.\,\ref{energy}. Typically, the kinetic power is given by the following relation, where $\mathrm{\dot M}$ is the mass loss rate and V$_{\infty}$ is the terminal velocity:
\begin{equation}
\mathrm{P_{kin}} = \frac{1}{2}\,{\mathrm{\dot M}}\,\mathrm{V_{\infty}}^2 .
\end{equation}
As a rule of thumb, an O-type star with a bolometric luminosity of the order of 5\,\,10$^{-6}$\,L$_\odot$, a mass loss rate of 10$^{-6}$\,M$_\odot$\,yr$^{-1}$, and a terminal velocity of 2000\,km\,s$^{-1}$, has a $\mathrm{P_{kin}}$ of the order of 10$^{-3}$ times the bolometric luminosity of the star.

The derived values for the kinetic power are given in the seventh column of Table\,\ref{param}. It is interesting to note that the $\mathrm{P_{kin}}$ values spread over slightly less than three orders of magnitude, such that the wind parameter space is large. It is also noteworthy that there is no strict discrimination between O-type and WR-type systems in this distribution of kinetic power, lending more support to the unified approach adopted in this study, even though the largest kinetic power are found mainly among WR systems because of their significantly enhanced mass loss rates with respect to their OB progenitors.

As we are dealing with binary systems, it may also be instructive to evaluate the coverage of the orbital periods, as a function, for instance, of the kinetic power of the dominant wind. This is illustrated in Fig.\,\ref{period}. Orbital periods are determined for only about 50 percent of the catalogue members. About 3 orders of magnitude in known orbital periods are covered by PACWBs, and it should extend to at least one more order of magnitude if the longer period systems (e.g. WR\,146 or WR\,147, whose orbital periods are not determined) are also included. The evidence that only a fraction of PACWBs have clearly identified orbital periods points to a severe bias in our knowledge of orbital parameters, especially as wider systems are considered. However, one can already point out that the capability to accelerate particles is not limited to a narrow range of orbital periods, lending support to the idea that particle acceleration processes can take place in a wide range of conditions. This suggests a priori that particle acceleration could be a common feature in colliding-wind binaries, even though many of the systems may be out of reach of most current observational facilities. 

\subsection{Energy budget}\label{energy}

\begin{figure*}[ht]
\begin{center}
\includegraphics[width=160mm]{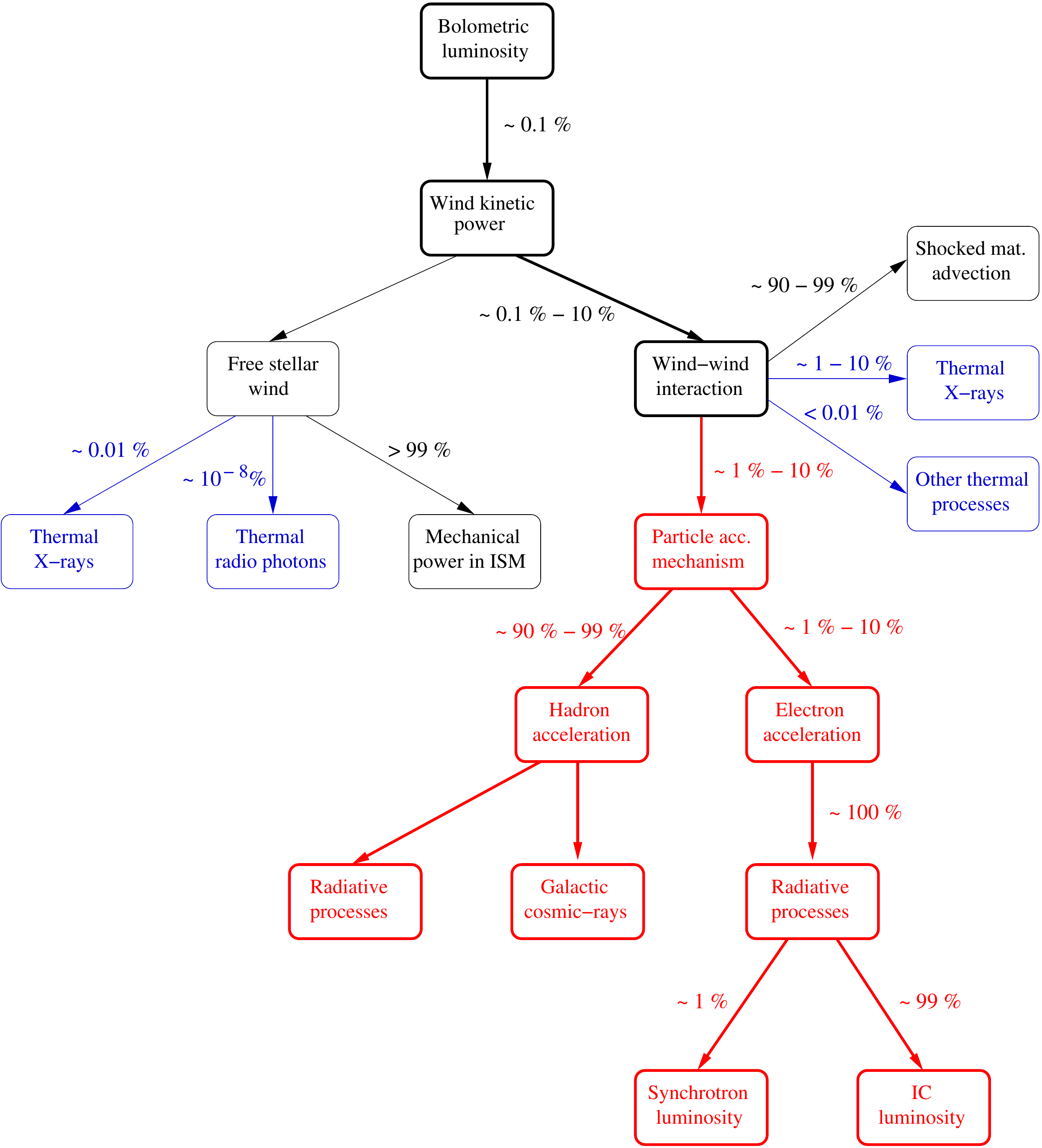}
\caption{Energy budget of stellar winds in a colliding-wind binary with emphasis on non-thermal processes (in red). Thermal processes are also mentioned (in blue). The energy transfer from the stellar bolometric luminosity to non-thermal emission processes in indicated by thick boxes and arrows. The numbers expressed in $\%$ provide orders of magnitude for the energy transfer between consecutive boxes. These numbers correspond to typical values likely to differ significantly from one object to the other.\label{enbud}}
\end{center}
\end{figure*}

The investigation of PACWBs offers the opportunity to study acceleration processes in massive binaries. One relevant consideration is that of the energy budget available for the particle acceleration mechanism, as illustrated in Fig.\,\ref{enbud}. The radiation field of a massive star will accelerate the wind material through the line-driving mechanism \citep{LS,CAK}. A low fraction of the bolometric luminosity is converted into kinetic power of the stellar wind. The efficiency of this transfer is dependent on the composition of the outer layers of the star. It is indeed well established that the line-driving mechanism of stellar winds is much more efficient in Wolf-Rayet stars, characterized by higher metallicities. This translates into generally significantly higher mass loss rates for WR stars as compared to O-type and B-type stars by at least one order of magnitude (see Fig.\,\ref{mdotvterm}). 

As one considers particle acceleration in the collision region, only the part of the wind kinetic power injected in the wind-wind interaction region is potentially available for the particle acceleration mechanism. In long period massive binaries, one may consider that the pre-shock velocity is the terminal velocity of the stellar wind, since the stagnation point of the colliding-winds will be located outside the acceleration zone of the stellar winds. Considering the fraction f of the outflowing plasma moving to the direction of the wind-wind interaction region, the part of the kinetic power that should be considered is
\begin{equation}
\mathrm{P_{ww}} = \frac{1}{2}\,\mathrm{f}\,{\mathrm{\dot M}}\,\mathrm{V_{\infty}}^2 .
\end{equation}
The factor f is typically a geometric dilution factor scaling with the fractional solid angle covered by the collision region as seen from the star's point of view. The complement of the fraction f (i.e., 1 -- f) will be predominantly released in the form of mechanical energy in the interstellar medium. 

Let us consider that the stars with the highest and lowest wind momentum rates (respectively, stars A and B) are located at distances r$_\mathrm{A}$ and r$_\mathrm{B}$, respectively, from the stagnation point. According to \citet{EU}, the extension of the non-thermal emitting region is of the order of $\pi$\,r$_\mathrm{B}$. The total projected surface exposed to the interaction with the stellar wind of star A, leading to non-thermal emission processes, is 
\begin{equation}
\mathrm{S} = \pi\,\big(\frac{\pi\,\mathrm{r_B}}{2}\big)^2 .
\end{equation}

This characteristic surface corresponds to a solid angle expressed in steradians (as seen from the star with the strongest wind momentum rate):
\begin{equation}
\Omega = \frac{S}{\mathrm{r}_A^2} = \frac{\pi^3}{4}\,\frac{\mathrm{r_B^2}}{\mathrm{r_A^2}} .
\end{equation}
We therefore derive a relation for the typical fractional solid angle (f)
\begin{equation}\label{factorf}
\mathrm{f} = \frac{\Omega}{4\,\pi} = \frac{\pi^2}{16}\,\eta ,
\end{equation}
where $\eta$ is the wind momentum rate ratio, as defined by \citet{SBP}. The value derived for the f factor, of course, depends intimately on specific circumstances related to the wind properties and orbital parameters of every object. However, one can see for a general discussion aiming at providing orders of magnitudes in energetic considerations that Eq.\,\ref{factorf} leads to values of the order of a fraction of a percent to several percents (as mentioned in Fig.\,\ref{enbud}) in cases where the wind momentum rate ratio gets closer to unity. One should note that the energy transfer to the wind-wind interaction region discussed here is independent on the size of the orbit. Systems with different orbital elements, but similar $\eta$ values transfer energy with similar efficiencies. However, systems with different sizes are characterized by different energy densities (i.e., energy per unit volume).

Depending on the wind-wind interaction region geometry, the effective injection of kinetic power spreads over a wide range of values. The pre-shock velocity vectors could be split into perpendicular and tangential components with relative values depending on the curvature of the interaction region and on the angular distance measured with respect to the line of centres. In an asymmetric system, the energy injection from the dominating wind is therefore maximum along that line and decreases significantly as the incidence angle increases. On the contrary, the same geometrical considerations lead to a much higher injection efficiency for the secondary wind, around which the shock cone is warped. The largest effective contribution to the injected kinetic power does not therefore come necessarily from the dominating wind. However, the current lack of information about the nature of the companion in a high fraction of the catalogue limits our capability to consider this effect adequately. Ideally, an effective kinetic power injection rate f$_\mathrm{eff}$ should be considered in practise.

In addition, a fraction $\kappa_\mathrm{acc}$ of the injected kinetic energy participates in the acceleration of particles, leading to the following expression for the amount of energy per unit time transferred to accelerated particles:
\begin{equation}
\mathrm{P_{acc}} = \frac{1}{2}\,\kappa_\mathrm{acc}\,\mathrm{f_{eff}}\,{\mathrm{\dot M}}\,\mathrm{V_{\infty}}^2 .
\end{equation}

The energy involved in the particle acceleration mechanism is shared into electron acceleration and hadron acceleration. Electrons are involved in synchrotron radio emission (Sect.\,\ref{radiosection}) and inverse Compton scattering responsible for non-thermal X-rays (Sect.\,\ref{highenergysection}). Accelerated hadrons consist mainly of protons, but helium and other elements nuclei are also expected to contribute. On the one hand, the census of Galactic cosmic rays suggest an electron-to-hadron ratio of the order of 0.01 \citep{Long}. Even though Galactic cosmic-rays are expected to come mainly from supernova remnants \citep{snrcr1,aharoniansnr} and not from colliding-wind massive binaries, one could a priori expect this ratio to be similar for astrophysical environments giving rise to a very similar acceleration process. On the other hand, values as high as 0.1 has been considered for instance by \citet{EU} in the case of WR\,140. As a first guess, we therefore consider that the efficiency ratio for the acceleration of electrons and hadrons ($\beta$) should lie between 0.01 and 0.1. Considering the $\beta$ parameter, the energy per unit time made available to the relativistic electron population can be written as:
\begin{equation}
\mathrm{P_{el}} = \frac{1}{2}\,\beta\,\kappa_\mathrm{acc}\,\mathrm{f_{eff}}\,{\mathrm{\dot M}}\,\mathrm{V_{\infty}}^2 .
\end{equation} 

Considering the wealth of UV and visible photons coming from the massive stars in the PACWB, relativistic electrons would most likely be thermalized through IC scattering with almost no chance to escape the wind-wind interaction region before losing the energy gained through the particle acceleration process. The ratio of synchrotron to IC scattering energy losses for electrons is that of the magnetic to photon energy densities \citep{PD140art,debeckerreview}. In typical massive binary conditions, this ratio is strongly in favor of IC scattering. As a result, almost 100\,$\%$ of $\mathrm{P_{el}}$ should be radiated in the form of non-thermal high energy photons. Only a low fraction ($\zeta$) of the amount of energy injected in relativistic electrons is converted into synchrotron radio emission. We can therefore write the following relation between synchrotron radio luminosity and the wind parameters:
\begin{equation}
\mathrm{L_{synch}} = \frac{1}{2}\,\zeta\,\beta\,\kappa_\mathrm{acc}\,\mathrm{f_{eff}}\,{\mathrm{\dot M}}\,\mathrm{V_{\infty}}^2 .
\end{equation}

These considerations clearly show that the available kinetic power plays a pivotal role in the production of synchrotron radio photons with a proportionality between these two quantities depending on several factors that might differ significantly from one system to the other: $\mathrm{L_{synch}} \propto \mathrm{P_{kin}}$. One may wonder what could be the dispersion of this proportionality relation. Here, one could define an intrinsic radio synchrotron efficiency factor (RSE), defined by the ratio $\mathrm{L_{synch}/P_{kin} = \zeta\,\beta\,\kappa_{acc}\,f_{eff}}$.

\begin{figure*}[ht]
\begin{center}
\includegraphics[width=140mm]{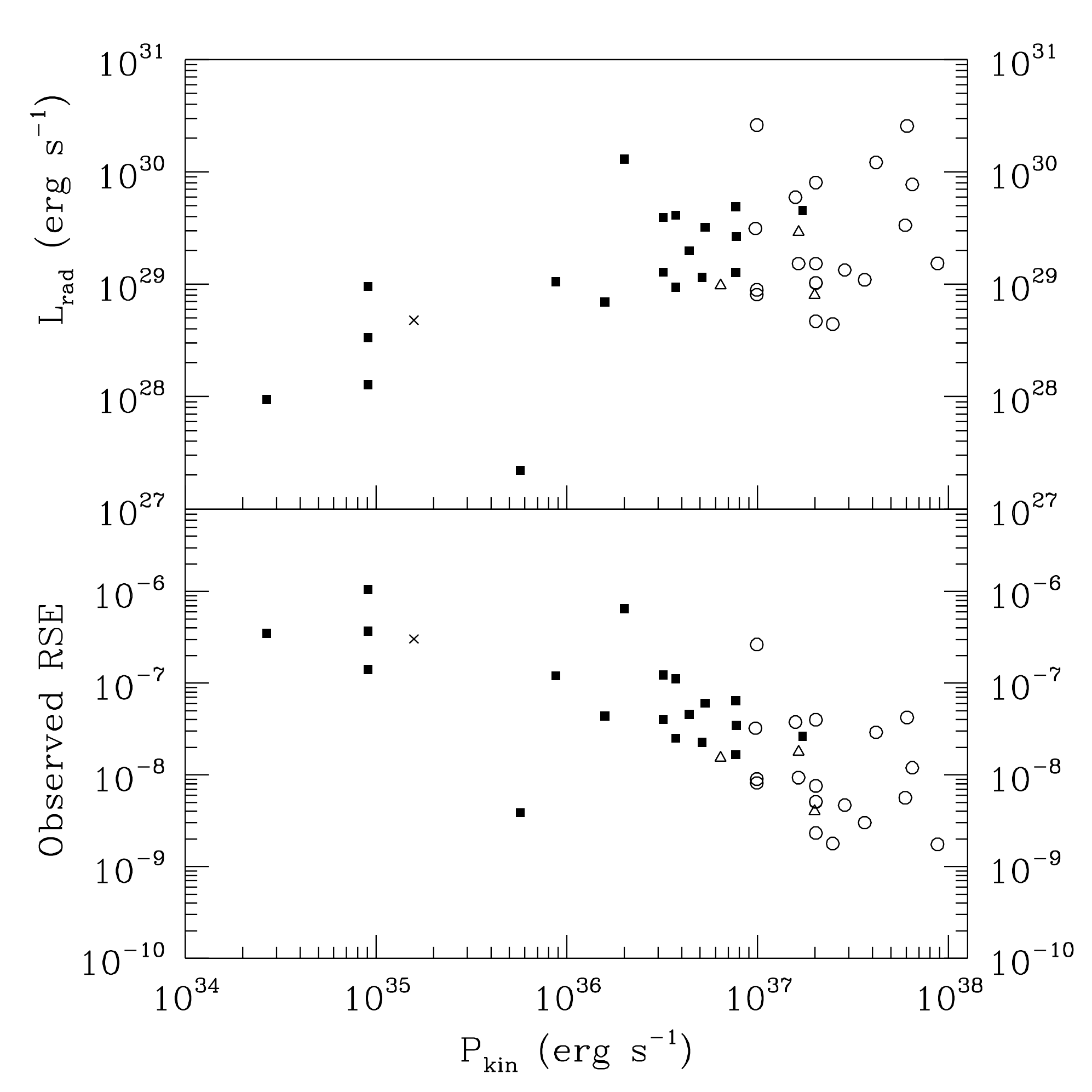}
\caption{Estimated radio luminosity ({\it Upper panel}) and ratio of the radio luminosity over the kinetic power of the dominant wind (RSE, {\it Lower panel}) as a function of the kinetic power of the dominant wind. The symbols have the same meaning as in Fig.\,\ref{mdotvterm}.\label{lratio}}
\end{center}
\end{figure*}

To have a rough idea of the radio synchrotron efficiency in PACWBs, the radio luminosity can be estimated on the basis of information collected in the literature and summarized in Table\,\ref{param}, using the relations given in \citet{debeckerreview}. It should be emphasized here that observations alone allow to estimate an observed RSE and not an intrinsic one. The measured flux densities are indeed affected by free-free absorption. In this approach, several criteria were adopted:
\begin{enumerate}
\item[-] Radio flux density measurements at a short wavelength (6\,cm, or even 3.6\,cm) were favored, as the free-free absorption is expected to be less pronounced than at longer wavelengths.
\item[-] For each object, the highest value was selected when several flux densities were reported in the literature, provided measurements exist at the same epoch and at other wavelengths to derive a spectral index. If the flux density is measured at only one wavelength at the epoch of maximum emission, another epoch was selected to be able to determine a spectral index. The decision to consider the highest recorded value of the flux density allows us to assume that the radio emission is dominated by synchrotron radiation (i.e., thermal contributions from the individual stellar winds are neglected).
\item[-] The spectral index given by the authors of the relevant studies was considered when available, and it was calculated on the basis of flux density measurements at two distinct wavelengths otherwise. When a value for the flux density was found only at one wavelength, an arbitrary spectral index equal to 0.0 was assumed.
\item[-] The observed radio luminosity for each object was calculated assuming distances given in Table\,\ref{cata} (except for $\eta$\,Car for which there is no detection of synchrotron radio emission and for WR\,137 for which a published value for the spectral index was found without any value for the flux density).
\end{enumerate}
The calculated values for the radio luminosities (L$_\mathrm{rad}$, expressed in erg\,s$^{-1}$) are given in the last column of Table\,\ref{param}. In the cases where no spectral index could be found or calculated from published quantities, L$_\mathrm{rad}$ is given in parentheses. To achieve an overview of the radio synchrotron efficiency for the members of the catalogue, we plotted L$_\mathrm{rad}$ and the observed RSE (\,=\,L$_\mathrm{rad}$/P$_\mathrm{kin}$) as a function of P$_\mathrm{kin}$ as shown in the lower panel of Fig.\,\ref{lratio} (in logarithm form for display purpose). One sees that RSE ranges between 10$^{-9}$ and 10$^{-6}$ with wind kinetic power spreading over values of a few 10$^{34}$\,erg\,s$^{-1}$ up to several 10$^{37}$\,erg\,s$^{-1}$. As could be anticipated on the basis of the plot of the wind parameters shown in Fig.\,\ref{mdotvterm}, most WR stars are located in the region of the plot corresponding to the highest kinetic powers. One should emphasize here that this plot could not be considered as an in-depth probe of the efficiency of non-thermal processes in massive binaries but rather as an indicator of some likely trends that deserve to be ascertained through future investigations. In particular, it is noticeable that the largest values for the observed RSE are preferentially found for lower kinetic powers. This apparent trend could at least be partly explained if one reminds that RSE values are computed for observed values of the radio flux density, therefore likely affected by free-free absorption. Stellar winds with lower kinetic power are also less opaque to radio photons, therefore leading to higher apparent synchrotron efficiencies. We caution also that this plot should be viewed as some kind of snapshot and not as a permanent census of the non-thermal radio energy budget of PACWBs. The synchrotron radio emission is highly variable on the orbital time scale, and in most cases, the radio emission has not been monitored adequately. In addition, the colliding-wind region in very long period systems is likely to be significantly active at accelerating particles, producing synchrotron emission, at orbital phases not too far from periastron. At this stage, such a plot has the benefit to provide a first representation of the location of PACWBs in the $\mathrm{P_{kin}}-\mathrm{L_{rad}}$ plane for the most complete population that has been identified so far with the information available so far.

\subsection{Open questions related to PACWBs}
In the context of this topic, it is obvious that several open questions remain. A non-exhaustive list is briefly commented below.

\paragraph{The multiplicity of a few systems.}
Some members of {\it list\,A} deserve to be investigated in either spectroscopy or interferometry to search for the presence of a companion. It should be also emphasized that a few members of {\it list\,A} are only strongly suspected binaries on the basis of indirect evidence, but they still deserve dedicated observations to clarify this issue. In addition, some systems are identified to be at least binary, which does not mean that the issue of their multiplicity is definitely solved. The presence of additional stars is expected in a few cases, and additional studies are needed to ascertain the issue of their multiplicity. 

\paragraph{The fraction of PACWBs among CWBs.}
One may wonder whether undetected PACWBs emit non-thermal radiation at a level below present instrument sensitivity or whether some specific circumstances are required for particle acceleration processes to operate in massive binaries. It is also striking that PACWBs cover a wide stellar, wind and orbital parameter space, and one may wonder why other systems located in the same part of the parameter space have not revealed any non-thermal signature yet. Is it coming from observational biases, or does it translate an actual lack of non-thermal activity in many systems? This issue is probably one of the most important ones in the context of particle acceleration in massive stars. 

\paragraph{The efficiency of particle acceleration.}
The issue of the efficiency of the energy transfer from interacting winds to accelerated particles in also very important. As discussed in Sect.\,\ref{energy}, at most a few percent of the energy involved in the colliding-winds is expected to be injected in relativistic particles. This issue is important in the sense that it might affect the hydrodynamics of colliding-winds (e.g. the shock modification scenario).

\paragraph{The electron-to-hadron ratio.}
The acceleration efficiency of electrons and protons, or other hadrons, such as alpha particles and orther nuclei, is also a key issue. Both radiative processes involving electrons (inverse Compton scattering and synchrotron radiation) and protons (neutral pion decay) need to be investigated in detail to quantify the amount of energy transferred to both types of particles.

\paragraph{The effect of wind composition.}
As illustrated in this catalogue, the parameter space covered by PACWBs is large, including very different wind compositions ranging between almost solar abundances for regular O-type stars to hydrogen depleted and carbon rich envelopes of WC-type stars. These differences in element abundances are a priori expected to lead to differences in the populations of relativistic particles accelerated in PACWBs. This issue has never been investigated so far.

\paragraph{The contribution of PACWBs to Galactic cosmic-rays.}
It is now well accepted that Galactic cosmic-rays are mainly accelerated in supernova remnants (SNRs). However, one may ask whether PACWBs could at least partly contribute to the Galactic cosmic-ray population. SNRs are known to be efficient at accelerating particles during the Sedov expansion phase, lasting for $\sim 10^4$\,yr. Even though the efficiency of PACWBs at accelerating particles is expected to be much lower than that of SNRs, they are able to accelerate particles for several 10$^6$\,yr. Integrating over the whole Galactic population of CWBs and SNRs, one cannot reject a priori the idea that the contribution from PACWBs may be significant (although weak) for low energy cosmic-rays (up to $\sim$\,1\,GeV). For very high energy cosmic-rays, it seems however unlikely that PACWBs could contribute, since their capability to accelerate particles should at most be limited to energies of a few GeV for electrons and, perhaps, a few 100\,GeV for protons. This issue is very important in the context of the production of Galactic cosmic-rays in general.

\section{Summary and conclusions}\label{conclusionssection}
This paper describes the first unified catalogue devoted to particle-accelerating colliding-wind binaries (PACWBs), covering spectral classifications from early B-type to evolved WR-stars. This catalogue includes a detailed census of the multiplicity of its members, along with relevant information related to radio and high-energy observations. This paper also includes information related to additional objects that claim to be particle accelerator candidates but deserve more studies before ascertaining their status. The publication of this paper is simultaneous with the operation of an on-line version of this catalogue, which is likely to be updated as new information will be made available in the future. The on-line version of the catalogue can be queried via this web-link: {\tt http://www.astro.ulg.ac.be/$\sim$debecker/pacwb}.

The unified approach followed in this study allows us to discuss the spread of the stellar and wind parameters of these objects, emphasizing the size of the parameter space allowing colliding-wind massive binaries to accelerate particles up to relativistic energies. The wide extension of the parameter space populated by PACWBs suggests that particle acceleration could be a rather common feature in colliding-wind binaries, even though many systems participating in these processes may produce a non-thermal activity well below the sensitivity of current observational facilities. The global view on this category of objects achieved through this catalogue is intended to provide clues for future observational and theoretical investigations aiming to clarify the physics of particle acceleration and non-thermal emission processes in massive binaries. In this context, a non-exhaustive list of open questions is enumerated, providing guidelines for future studies.

\begin{acknowledgements}
The author wants to express his gratitude to Drs. Julian Pittard, Hugues Sana, Ronny Blomme, Gustavo E. Romero, Dmitry Khangulyan, Valenti Bosch-Ramon and Eric Gosset for fruitful collaborations and/or stimulating discussions over the past few years on different aspects of the science topic addressed in this paper. We also thank the anonymous referee for a detailed reading of the paper and for helpful and constructive comments that improved the manuscript. The SIMBAD database was used for the bibliography.
\end{acknowledgements}

\bibliographystyle{aa}
\bibliographystyle{aa}
%\bibliography{bibmd}


\newpage

\appendix

\section{The catalogue}

\setcounter{id}{0}
\newcounter{refd}
\begin{table*}[h]
\caption{Catalogue of particle-accelerating colliding-wind binaries, sorted by increasing right ascension.\label{cata}}
\begin{center}
\begin{tabular}{l l l l l l l l}
\hline
\# & usual ID & other ID & $\alpha$[J2000] & $\delta$[J2000] & d (pc) & Ref. & Ass./Clu. \\
\hline
\vspace*{-0.2cm}\\
\stepcounter{id}\theid & \object{HD~15558} & BD~+60\,502 & 02 32 42.54 & +61 27 21.58 & 2300 & \stepcounter{refd}\therefd &  IC\,1805 \\
\stepcounter{id}\theid & $\delta$~Ori\,A & \object{HD~36486} & 05 32 00.40 & --00 17 56.74 & 360,473 & \stepcounter{refd}\therefd,\stepcounter{refd}\therefd & Ori\,OB1b \\
\stepcounter{id}\theid & $\sigma$~Ori\,AB & \object{HD~37468} & 05 38 44.78 & --02 36 00.12 & 440 & \stepcounter{refd}\therefd & $\sigma$\,Ori \\
\stepcounter{id}\theid & 15\,Mon & \object{HD~47839} & 06 40 58.66 & +09 53 44.72 & 950 & \stepcounter{refd}\therefd & Mon\,OB1 \\
\stepcounter{id}\theid & WR\,8 & \object{HD~62910} & 07 44 58.22 & --31 54 29.55 & 3470 & \stepcounter{refd}\therefd \newcounter{vdhd} \setcounter{vdhd}{\value{refd}} & -- \\
\stepcounter{id}\theid & \object{WR~11} & $\gamma^2$\,Vel & 08 09 31.95 & --47 20 11.71 & 350 & \stepcounter{refd}\therefd,\stepcounter{refd}\therefd &  Vel\,OB2 \\
\stepcounter{id}\theid & \object{WR~14} & HD\,76536 & 08 54 59.17 & --47 35 32.68 & 2000(*) & \stepcounter{refd}\therefd & -- \\
\stepcounter{id}\theid & \object{CD--47\,4551} & CPD--47\,2963 & 08 57 54.62 & --47 44 15.73 & 1300 & \stepcounter{refd}\therefd\newcounter{bena} \setcounter{bena}{\value{refd}}  & --\\
\stepcounter{id}\theid & \object{WR~21a} & Th35-42 & 10 25 56.50 & --57 48 43.5 & 3000 & \stepcounter{refd}\therefd & -- \\
\stepcounter{id}\theid & \object{HD~93129A} & CD--58\,3527 & 10 43 57.40 & --59 32 52.31 & 2000(*) & \stepcounter{refd}\therefd & Tr\,14 \\
\stepcounter{id}\theid & \object{HD~93250} & CD--58\,3537 & 10 44 45.03 & --59 33 54.68 & 2350 & \stepcounter{refd}\therefd  \newcounter{smith} \setcounter{smith}{\value{refd}}& Tr\,16 \\
\stepcounter{id}\theid & $\eta$\,Car & \object{HD~93308} & 10 45 03.59 & --59 41 04.26 & 2350 & \thesmith & Tr\,16 \\
\stepcounter{id}\theid & \object{WR~39} & MS\,9 & 11 06 18.70 & --61 14 18.3 & 5700 & \stepcounter{refd}\therefd \newcounter{smithwr} \setcounter{smithwr}{\value{refd}} & -- \\
\stepcounter{id}\theid & \object{WR~48} & $\theta$\,Mus & 13 08 07.15 & --65 18 21.68 & 2400 & \stepcounter{refd}\therefd & Cen\,OB1 \\
\stepcounter{id}\theid & \object{HD~124314} & CD~--61\,4297 & 14 15 01.61 & --61 42 24.38 & 1000 & \thebena & -- \\
\stepcounter{id}\theid & \object{HD~150136} & HR\,6187 & 16 41 20.42 & --48 45 46.75 & 1320 & \stepcounter{refd}\therefd & Ara\,OB1 \\
\stepcounter{id}\theid & \object{HD~151804} & CD~--41\,10957 & 16 51 33.72 & --41 13 49.93 & 1675(**) & \stepcounter{refd}\therefd & Sco\,OB1 \\
\stepcounter{id}\theid & \object{WR~78} & HD\,151932 & 16 52 19.25 & --41 51 16.26 & 1600 &\stepcounter{refd}\therefd \newcounter{scoob} \setcounter{scoob}{\value{refd}}  & Sco\,OB1 \\
\stepcounter{id}\theid & WR\,79a & \object{HD~152408} & 16 54 58.51 & --41 09 03.10 & 1600 & \thescoob & Sco\,OB1 \\
\stepcounter{id}\theid & \object{HD~152623} & CD~--40\,10961 & 16 56 15.03 & --40 39 35.81 & 1600 & \thescoob & Sco\,OB1 \\
\stepcounter{id}\theid & \object{WR~89} & CD~--38\,11746 & 17 19 00.52 & --38 48 51.25 & 3300 & \stepcounter{refd}\therefd & HM\,1 \\
\stepcounter{id}\theid & \object{WR~90} & HD\,156385 & 17 19 29.90 & --45 38 23.88 & 1640 & \thevdhd & -- \\
\stepcounter{id}\theid & \object{WR~98} & HD\,318016 & 17 37 13.75 & --33 27 55.98 & 1900 & \stepcounter{refd}\therefd & -- \\
\stepcounter{id}\theid & \object{WR~98a} & IRAS\,17380--3031 & 17 41 12.9 & --30 32 29 & 1900 & \stepcounter{refd}\therefd & -- \\
\stepcounter{id}\theid & \object{WR~104} & Ve2-45 & 18 02 04.13 & --23 37 42.0 & 1600 & \stepcounter{refd}\therefd & Sgr\,OB1 \\
\stepcounter{id}\theid & \object{WR~105} & Ve2-47 & 18 02 23.45 & --23 34 37.5 & 1580 & \thevdhd & Sgr\,OB1 \\
\stepcounter{id}\theid & 9\,Sgr & \object{HD~164794} & 18 03 52.45 & --24 21 38.63 & 1580 & \stepcounter{refd}\therefd & NGC\,6530 \\
\stepcounter{id}\theid & \object{WR~112} & GL\,2104 & 18 16 33.49 & --18 58 42.3 & 4150 & \thevdhd & -- \\
\stepcounter{id}\theid & \object{HD~167971} & BD--12\,4980 & 18 18 05.89 & --12 14 33.29 & 1700 & \stepcounter{refd}\therefd \newcounter{serob} \setcounter{serob}{\value{refd}} & NGC\,6604 \\
\stepcounter{id}\theid & \object{HD~168112} & BD--12\,4988 & 18 18 40.87 & --12 06 23.37 & 1700 & \theserob & NGC\,6604 \\
\stepcounter{id}\theid & CEN1a & \object{Kleinmann's Star} a & 18 20 29.90 & --16 10 44.40 & 2100 & \stepcounter{refd}\therefd\newcounter{cend} \setcounter{cend}{\value{refd}} & NGC\,6618 \\
\stepcounter{id}\theid & CEN1b & Kleinmann's Star b & 18 20 29.81 & --16 10 45.67 & 2100 & \thecend & NGC\,6618 \\
\stepcounter{id}\theid & \object{WR~125} & V378\,Vul & 19 28 15.62 & +19 33 21.4 & 1990 & \thesmithwr & -- \\
\stepcounter{id}\theid & \object{HD~190603} & BDv+31\,3925 & 20 04 36.18 & +32 13 06.95 & 1500 & \stepcounter{refd}\therefd,\stepcounter{refd}\therefd & -- \\
\stepcounter{id}\theid & \object{WR~133} & HD\,190918 & 20 05 57.32 & +35 47 18.15 & 2140 & \thevdhd & NGC\,6871 \\
\stepcounter{id}\theid & \object{WRv137} & HD\,192641 & 20 14 31.77 & +36 39 39.60 & 1700 & \stepcounter{refd}\therefd \newcounter{cygob} \setcounter{cygob}{\value{refd}} & Cyg\,OB1 \\
\stepcounter{id}\theid & \object{WR~140} & HD\,193793 & 20 20 27.98 & +43 51 16.28 & 1810 & \stepcounter{refd}\therefd & -- \\
\stepcounter{id}\theid & Cyg\,OB2\,\#5 & \object{BD~+40 4220} & 20 32 22.42 & +41 18 18.96 & 1700 & \thecygob & Cyg\,OB2 \\
\stepcounter{id}\theid & Cyg\,OB2\,\#9 & \object{Schulte 9} & 20 33 10.74 & +41 15 08.21 & 1700 & \thecygob & Cyg\,OB2 \\
\stepcounter{id}\theid & Cyg\,OB2\,\#8A & \object{BD+40 4227} & 20 33 15.08 & +41 18 50.48 & 1700 & \thecygob &  Cyg\,OB2 \\
\stepcounter{id}\theid & Cyg\,OB2-335 & \object{MT91 771} & 20 34 29.60 & +41 31 45.54 & 1700 & \thecygob & Cyg\,OB2 \\
\stepcounter{id}\theid & \object{WR~146} & HM~19-3 & 20 35 47.08 & +41 22 44.6 & 1200 & \stepcounter{refd}\therefd & -- \\
\stepcounter{id}\theid & \object{WR~147} & AS\,431 & 20 36 43.64 & +40 21 07.6 & 650 & \stepcounter{refd}\therefd & -- \\
\vspace*{-0.2cm}\\
\hline
\end{tabular}
\end{center}
\footnotesize
(*) value based on published distance modulus; (**) value determined from Hipparcos parallax.\\
\tablebib{
\setcounter{refd}{0}
\stepcounter{refd}\therefd. \citet{massey1805}; \stepcounter{refd}\therefd. \citet{brownoriob1}; \stepcounter{refd}\therefd. \citet{dezeeuworiob1}; \stepcounter{refd}\therefd. \citet{sherrysigori}; \stepcounter{refd}\therefd. \citet{15monperez}; \stepcounter{refd}\therefd. \citet{vdhcata}; \stepcounter{refd}\therefd. \citet{gam2velorbsol}; \stepcounter{refd}\therefd. \citet{millourwr11}; \stepcounter{refd}\therefd. \citet{wr14crowther}; \stepcounter{refd}\therefd. \citet{Benaglia2006}; s\stepcounter{refd}\therefd. \citet{BRKP}; \stepcounter{refd}\therefd. \citet{Walborn1995}; \stepcounter{refd}\therefd. \citet{smithcarinae}; \stepcounter{refd}\therefd. \citet{smith1990}; \stepcounter{refd}\therefd. \citet{Chap}; \stepcounter{refd}\therefd. \citet{HH1977}; \stepcounter{refd}\therefd. \citet{hns2010}; \stepcounter{refd}\therefd. \citet{reipurth6231}; \stepcounter{refd}\therefd. \citet{HM1dist}; \stepcounter{refd}\therefd. \citet{montes2009}; \stepcounter{refd}\therefd. \citet{pwnwr98a}; \stepcounter{refd}\therefd. \citet{wr104dist}; \stepcounter{refd}\therefd. \citet{9sgrdist}; \stepcounter{refd}\therefd. \citet{reipurth6604}; \stepcounter{refd}\therefd. \citet{hoffmeister2008}; \stepcounter{refd}\therefd. \citet{hd190603I}; \stepcounter{refd}\therefd. \citet{hd190603II}; \stepcounter{refd}\therefd. \citet{reipurthcyg}; \stepcounter{refd}\therefd. \citet{DougLIAC}; \stepcounter{refd}\therefd. \citet{Doug146dist}; \stepcounter{refd}\therefd. \citet{wr147morris}.
}
\end{table*}
\normalsize

\newpage

\setcounter{id}{0}
\newcounter{refmulti}
\begin{table*}[ht]
\caption{Multiplicity and spectral types of PACWBs.\label{multi}}
\begin{center}
\begin{tabular}{l l l l l l}
\hline
\# & usual ID & Status & Sp. type(s) & P & Ref.\\
\hline
\vspace*{-0.2cm}\\
\stepcounter{id}\theid & HD\,15558 & B (T?) & O5.5III(f) + O7V & 442\,d & \stepcounter{refmulti}\therefmulti,\stepcounter{refmulti}\therefmulti \\
\stepcounter{id}\theid & $\delta$Ori\,A & T & (O9.5II + B0.5III) + B? & 5.733\,d/$>$\,100\,yr & \stepcounter{refmulti}\therefmulti,\stepcounter{refmulti}\therefmulti \\
\stepcounter{id}\theid & $\sigma$Ori\,AB & M & O9.5V + B0.5V (+ OBs?) & ? & \stepcounter{refmulti}\therefmulti \\
\stepcounter{id}\theid & 15\,Mon & B & O7V(f) + O9.5Vn & 25.3\,yr & \stepcounter{refmulti}\therefmulti \\
\stepcounter{id}\theid & WR\,8 & B & WN7 + WC(?) & 38\,d , 115\,d & \stepcounter{refmulti}\therefmulti,\stepcounter{refmulti}\therefmulti \\
\stepcounter{id}\theid & WR\,11 & B & WC8 + O7.5 & 78.53\,d & \stepcounter{refmulti}\therefmulti,\stepcounter{refmulti}\therefmulti \\
\stepcounter{id}\theid & WR\,14 & B? & WC7 & ? & \stepcounter{refmulti}\therefmulti \newcounter{vdh} \setcounter{vdh}{\value{refmulti}} \\
\stepcounter{id}\theid & CD--47\,4551 & U & O5If & -- & \stepcounter{refmulti}\therefmulti \newcounter{benartsp} \setcounter{benartsp}{\value{refmulti}} \\
\stepcounter{id}\theid & WR\,21a & B & O3f$^*$/WN6ha + O4 & 32.673\,d & \stepcounter{refmulti}\therefmulti \\
\stepcounter{id}\theid & HD\,93129A & B & O2If$^*$ + O3.5V? & ? & \stepcounter{refmulti}\therefmulti \\
\stepcounter{id}\theid & HD\,93250 & B & O4III + O4III & $>$\,100\,d & \stepcounter{refmulti}\therefmulti \\
\stepcounter{id}\theid & $\eta$\,Car & B & ? + ? & 2022.7\,d & \stepcounter{refmulti}\therefmulti \\
\stepcounter{id}\theid & WR\,39 & B? & WC7 & ? & \stepcounter{refmulti}\therefmulti  \\
\stepcounter{id}\theid & WR\,48 & T & (WC5 + O6-7V) + OI? & 19.138\,d/? & \stepcounter{refmulti}\therefmulti,\stepcounter{refmulti}\therefmulti \\
\stepcounter{id}\theid & HD\,124314 & B? & O6V(n)((f)) & ? & \stepcounter{refmulti}\therefmulti,\stepcounter{refmulti}\therefmulti \\
\stepcounter{id}\theid & HD\,150136 & T & (O3-3.5V((f$^+$)) + O5.5-6V((f))) + O6.5-7V((f)) & 2.675\,d/8,2\,yr & \stepcounter{refmulti}\therefmulti,\stepcounter{refmulti}\therefmulti,\stepcounter{refmulti}\therefmulti \\
\stepcounter{id}\theid & HD\,151804 & U & O8Iaf  & -- & \thebenartsp \\
\stepcounter{id}\theid & WR\,78 & U & WN7h  & -- & \thevdh \\
\stepcounter{id}\theid & WR\,79a & B & WN9ha + ? & many years & \stepcounter{refmulti}\therefmulti\newcounter{mason} \setcounter{mason}{\value{refmulti}} \\
\stepcounter{id}\theid & HD\,152623 & T & (O7V((f)) + OB?) + OB? & 3.9\,d/? & \themason \\
\stepcounter{id}\theid & WR\,89 & B & WN8h + OB & ? & \thevdh \\
\stepcounter{id}\theid & WR\,90 & U & WC7 & -- & \thevdh \\
\stepcounter{id}\theid & WR\,98 & B & WN7/WC + O8-9 & 48.7\,d & \stepcounter{refmulti}\therefmulti \\
\stepcounter{id}\theid & WR\,98a & B & WC9 + OB? & 565\,d & \stepcounter{refmulti}\therefmulti \\
\stepcounter{id}\theid & WR\,104 & B & WC9 + B0.5V & 220\,d & \stepcounter{refmulti}\therefmulti \\
\stepcounter{id}\theid & WR\,105 & U & WN9h & -- & \thevdh \\
\stepcounter{id}\theid & 9\,Sgr & B & O3.5V((f$^+$)) + O5V & $\sim$\,8.6\,yr & \stepcounter{refmulti}\therefmulti,\stepcounter{refmulti}\therefmulti \\
\stepcounter{id}\theid & WR\,112 & B? & WC9 + ? & ? & \stepcounter{refmulti}\therefmulti \\
\stepcounter{id}\theid & HD\,167971 & T & (O6-7V + O6-7V) + O8I & 3.321\,d/$\sim$\,20\,yr & \stepcounter{refmulti}\therefmulti,\stepcounter{refmulti}\therefmulti \\
\stepcounter{id}\theid & HD\,168112 & B? & O5.5III(f$^+$) (+ OB?) & $>$\,1\,yr & \stepcounter{refmulti}\therefmulti,\stepcounter{refmulti}\therefmulti \\
\stepcounter{id}\theid & CEN\,1a & B & O4 + ? & ? & \stepcounter{refmulti}\therefmulti\newcounter{cenmul} \setcounter{cenmul}{\value{refmulti}} \\
\stepcounter{id}\theid & CEN\,1b & B & O4 + ? & ? & \thecenmul \\
\stepcounter{id}\theid & WR\,125 & B & WC7 + O9III & $>$\,15\,yr,$\sim$\,20-22\,yr & \stepcounter{refmulti}\therefmulti,\stepcounter{refmulti}\therefmulti \\
\stepcounter{id}\theid & HD\,190603 & U & B1.5Ia & -- & \thebenartsp \\
\stepcounter{id}\theid & WR\,133 & B & WN5 + O9I & 112.4\,d & \stepcounter{refmulti}\therefmulti \\
\stepcounter{id}\theid & WR\,137 & B & WC7 + O9V-III & 13.05\,yr & \stepcounter{refmulti}\therefmulti \\
\stepcounter{id}\theid & WR\,140 & B & WC7 + O5 & 7.9\,yr & \stepcounter{refmulti}\therefmulti,\stepcounter{refmulti}\therefmulti \\
\stepcounter{id}\theid & Cyg\,OB2\,\#5 & Q & (Ofpe/WN9 + O6–7Ia) + OB? + B0V & 6.598\,d/6.7\,yr/$>$\,9000\,yr & \stepcounter{refmulti}\therefmulti,\stepcounter{refmulti}\therefmulti,\stepcounter{refmulti}\therefmulti \\
\stepcounter{id}\theid & Cyg\,OB2\,\#9 & B & O5I + O6-7I & 2.35\,yr & \stepcounter{refmulti}\therefmulti,\stepcounter{refmulti}\therefmulti,\stepcounter{refmulti}\therefmulti \\
\stepcounter{id}\theid & Cyg\,OB2\,\#8A & B & O6If + O5.5III(f) & 21.908\,d & \stepcounter{refmulti}\therefmulti \\
\stepcounter{id}\theid & Cyg\,OB2-335 & B & O7V + O9V & a few days(?) & \stepcounter{refmulti}\therefmulti \\
\stepcounter{id}\theid & WR\,146 & B (T?) & WC6 + O8? & many years ? & \stepcounter{refmulti}\therefmulti,\stepcounter{refmulti}\therefmulti \\
\stepcounter{id}\theid & WR\,147 & B & WN8 + B0.5V & many years ? & \stepcounter{refmulti}\therefmulti \\
\vspace*{-0.2cm}\\
\hline
\end{tabular}
\end{center}
\setcounter{refmulti}{0}
\footnotesize{
Meaning of the status symbols: B is for binary system; T for triple system; Q: for quadruple system; B? for suggested binary; T?: for suggested triple system; M for multiple with undetermined number of components; U for undetermined.\\
\tablebib{\setcounter{refmulti}{0}
\stepcounter{refmulti}\therefmulti. \citet{GM}; \stepcounter{refmulti}\therefmulti. \citet{ic1805_2}; \stepcounter{refmulti}\therefmulti. \citet{deloriclosebin}; \stepcounter{refmulti}\therefmulti. \citet{deloritriple}; \stepcounter{refmulti}\therefmulti. \citet{drake}; \stepcounter{refmulti}\therefmulti. \citet{gies15mon}; \stepcounter{refmulti}\therefmulti. \citet{niemela1991binwr}; \stepcounter{refmulti}\therefmulti. \citet{wr8bin1998}; \stepcounter{refmulti}\therefmulti. \citet{gam2velorbsol1}; \stepcounter{refmulti}\therefmulti. \citet{gam2velorbsol}; \stepcounter{refmulti}\therefmulti. \citet{vdhcata}; \stepcounter{refmulti}\therefmulti. \citet{benagliareview}, and references therein; \stepcounter{refmulti}\therefmulti. \citet{wr21aniemela}; \stepcounter{refmulti}\therefmulti. \citet{nelan}; \stepcounter{refmulti}\therefmulti. \citet{hd93250vlti}; \stepcounter{refmulti}\therefmulti. \citet{etacarbin}; \stepcounter{refmulti}\therefmulti. \citet{Chap}; \stepcounter{refmulti}\therefmulti. \citet{wr48speck}; \stepcounter{refmulti}\therefmulti. \citet{wr48bin}; \stepcounter{refmulti}\therefmulti. \citet{feasthd124314}; \stepcounter{refmulti}\therefmulti. \citet{WDScata}; \stepcounter{refmulti}\therefmulti. \citet{NG150136}; \stepcounter{refmulti}\therefmulti. \citet{mahyhd150136}; \stepcounter{refmulti}\therefmulti. \citet{sana150136}; \stepcounter{refmulti}\therefmulti. \citet{masonspeckle}; \stepcounter{refmulti}\therefmulti. \citet{GN2003wr98}; \stepcounter{refmulti}\therefmulti. \citet{pwnwr98a}; \stepcounter{refmulti}\therefmulti. \citet{pwnwr104}; \stepcounter{refmulti}\therefmulti. \citet{multjenam}; \stepcounter{refmulti}\therefmulti. \citet{rauw9sgr2}; \stepcounter{refmulti}\therefmulti. \citet{marchenkowr112}; \stepcounter{refmulti}\therefmulti. \citet{Lei87}; \stepcounter{refmulti}\therefmulti. \citet{vlti167971}; \stepcounter{refmulti}\therefmulti. \citet{DeB168112}; \stepcounter{refmulti}\therefmulti. \citet{Blo168112}; \stepcounter{refmulti}\therefmulti. \citet{hoffmeister2008}; \stepcounter{refmulti}\therefmulti. \citet{DW}; \stepcounter{refmulti}\therefmulti. \citet{Cap}; \stepcounter{refmulti}\therefmulti. \citet{UH1994wr133}; \stepcounter{refmulti}\therefmulti. \citet{wr137orbit}; \stepcounter{refmulti}\therefmulti. \citet{marchenko140}; \stepcounter{refmulti}\therefmulti. \citet{fahed140mnras}; \stepcounter{refmulti}\therefmulti. \citet{rauwcyg5}; \stepcounter{refmulti}\therefmulti. \citet{kennedycyg5}; \stepcounter{refmulti}\therefmulti. \citet{Lindercyg5}; \stepcounter{refmulti}\therefmulti. \citet{vanloocyg9}; \stepcounter{refmulti}\therefmulti. \citet{nazecyg9}; \stepcounter{refmulti}\therefmulti. \citet{Blommecyg9}; \stepcounter{refmulti}\therefmulti. \citet{Let8a}; \stepcounter{refmulti}\therefmulti. \citet{kiminkicygob2}; \stepcounter{refmulti}\therefmulti. \citet{wr146setia}; \stepcounter{refmulti}\therefmulti. \citet{wr146companion}; \stepcounter{refmulti}\therefmulti. \citet{williamswr147}.}
}
\end{table*}
\normalsize

\newpage

\setcounter{id}{0}
\newcounter{refrad}
\begin{table*}[h]
\caption{Radio information on PACWBs.\label{radio}}
\begin{center}
\begin{tabular}{l l l l l}
\hline
\# & usual ID & Non-thermal (NT) criteria & Ref. & Q \\
\hline
\vspace*{-0.2cm}\\
\stepcounter{id}\theid & HD\,15558 & variable flux & \stepcounter{refrad}\therefrad \newcounter{bacart} \setcounter{bacart}{\value{refrad}} & II \\
\stepcounter{id}\theid & $\delta$\,Ori\,A & variable flux & \thebacart & II \\
\stepcounter{id}\theid & $\sigma$\,Ori\,AB & $\alpha < 0$ or close to 0 & \stepcounter{refrad}\therefrad \newcounter{drakeart} \setcounter{drakeart}{\value{refrad}} & I \\
\stepcounter{id}\theid & 15\,Mon & $\alpha < 0$, variable flux & \thedrakeart & I \\
\stepcounter{id}\theid & WR\,8 & $\alpha < 0.3$ & \stepcounter{refrad}\therefrad\newcounter{montesart}\setcounter{montesart}{\value{refrad}} & II \\
\stepcounter{id}\theid & WR\,11 & $\alpha \sim 0.3$ & \stepcounter{refrad}\therefrad\newcounter{leibart} \setcounter{leibart}{\value{refrad}},\stepcounter{refrad}\therefrad \newcounter{chapart} \setcounter{chapart}{\value{refrad}} & II \\
\stepcounter{id}\theid & WR\,14 & $\alpha < 0$, variable flux & \theleibart,\thechapart & I \\
\stepcounter{id}\theid & CD--47\,4551 & high flux density, $\alpha < 0$ & \stepcounter{refrad}\therefrad\newcounter{benart}\setcounter{benart}{\value{refrad}},\stepcounter{refrad}\therefrad\newcounter{bensaca}\setcounter{bensaca}{\value{refrad}},\stepcounter{refrad}\therefrad \newcounter{bckart}\setcounter{bckart}{\value{refrad}} & I \\
\stepcounter{id}\theid & WR\,21a & $\alpha < 0.3$ & \stepcounter{refrad}\therefrad & II \\
\stepcounter{id}\theid & HD\,93129A & high flux density, $\alpha < 0$, resolved NT emitting region & \stepcounter{refrad}\therefrad,\thebenart,\thebensaca,\stepcounter{refrad}\therefrad & I \\
\stepcounter{id}\theid & HD\,93250 & high flux density & \stepcounter{refrad}\therefrad & II \\
\stepcounter{id}\theid & $\eta$\,Car & -- & -- & I \\
\stepcounter{id}\theid & WR\,39 & $\alpha$ close to 0, offset of about 3'' w.r.t optical source & \thechapart & I \\
\stepcounter{id}\theid & WR\,48 & $\alpha < 0$ & \thechapart & I \\
\stepcounter{id}\theid & HD\,124314 & high flux density, $\alpha < 0$ & \thebenart,\thebensaca,\thebckart & I \\
\stepcounter{id}\theid & HD\,150136 & high flux density, $\alpha < 0$ & \thebenart,\thebensaca,\thebckart & I \\
\stepcounter{id}\theid & HD\,151804 & $\alpha < 0$ & \stepcounter{refrad}\therefrad \newcounter{scoart}\setcounter{scoart}{\value{refrad}} & I \\
\stepcounter{id}\theid & WR\,78 & $\alpha < 0$ & \thescoart & I \\
\stepcounter{id}\theid & WR\,79a & $\alpha < 0$, variable flux & \thescoart,\stepcounter{refrad}\therefrad \newcounter{capart} \setcounter{capart}{\value{refrad}} & I \\
\stepcounter{id}\theid & HD\,152623 & $\alpha < 0$ & \thescoart & I \\
\stepcounter{id}\theid & WR\,89 & variable flux & \thecapart & II \\
\stepcounter{id}\theid & WR\,90 & $\alpha$ close to 0, variable flux & \thechapart & I \\
\stepcounter{id}\theid & WR\,98 & $\alpha < 0.3$, variable flux & \themontesart & I \\
\stepcounter{id}\theid & WR\,98a & variable flux, $\alpha$ close to 0 & \thecapart,\stepcounter{refrad}\therefrad \newcounter{monart} \setcounter{monart}{\value{refrad}} & I \\
\stepcounter{id}\theid & WR\,104 & variable flux, $\alpha$ close to 0 & \thecapart,\themonart & I \\
\stepcounter{id}\theid & WR\,105 & variable flux, $\alpha < 0$ & \theleibart,\thechapart,\thecapart & I \\
\stepcounter{id}\theid & 9\,Sgr & high flux density, $\alpha < 0$ or close to 0, variable flux & \thebacart,\stepcounter{refrad}\therefrad & I \\
\stepcounter{id}\theid & WR\,112 & $\alpha$ close to 0, variable flux & \thechapart,\themonart & I \\
\stepcounter{id}\theid & HD\,167971 & high flux density, $\alpha < 0$, variable flux & \thebacart,\stepcounter{refrad}\therefrad & I \\
\stepcounter{id}\theid & HD\,168112 & high flux density, $\alpha < 0$, variable flux & \thebacart,\stepcounter{refrad}\therefrad,\stepcounter{refrad}\therefrad & I \\
\stepcounter{id}\theid & CEN\,1a & high flux density, variable flux and $\alpha$ close to 0 & \stepcounter{refrad}\therefrad\newcounter{cenrad}\setcounter{cenrad}{\value{refrad}},\stepcounter{refrad}\therefrad\newcounter{cenradnew} \setcounter{cenradnew}{\value{refrad}} & I \\
\stepcounter{id}\theid & CEN\,1b & high flux density, $\alpha < 0$ & \thecenrad,\thecenradnew & I \\
\stepcounter{id}\theid & WR\,125 & $\alpha < 0$, variable flux & \thechapart,\stepcounter{refrad}\therefrad,\stepcounter{refrad}\therefrad & I \\
\stepcounter{id}\theid & HD\,190603 & $\alpha < 0.3$ & \stepcounter{refrad}\therefrad & II \\
\stepcounter{id}\theid & WR\,133 & $\alpha < 0$ & \themontesart & I \\
\stepcounter{id}\theid & WR\,137 & $\alpha$ close to 0 & \stepcounter{refrad}\therefrad\newcounter{dwart}\setcounter{dwart}{\value{refrad}} & II \\
\stepcounter{id}\theid & WR\,140 & high flux density, $\alpha < 0$ and  variable, variable flux, resolved NT emitting region & \stepcounter{refrad}\therefrad,\stepcounter{refrad}\therefrad,\stepcounter{refrad}\therefrad,\stepcounter{refrad}\therefrad & I \\
\stepcounter{id}\theid & Cyg\,OB2\,\#5 & $\alpha < 0$, variable flux, resolved NT emitting region & \stepcounter{refrad}\therefrad,\stepcounter{refrad}\therefrad,\stepcounter{refrad}\therefrad & I \\
\stepcounter{id}\theid & Cyg\,OB2\,\#9 & $\alpha < 0$, variable flux, bow-shaped NT emission & \thebacart,\stepcounter{refrad}\therefrad,\stepcounter{refrad}\therefrad,\stepcounter{refrad}\therefrad & I \\
\stepcounter{id}\theid & Cyg\,OB2\,\#8A & $\alpha < 0$, variable flux & \thebacart,\stepcounter{refrad}\therefrad & I \\
\stepcounter{id}\theid & Cyg\,OB2-335 & high flux density, $\alpha < 0$ & \stepcounter{refrad}\therefrad\newcounter{wester} \setcounter{wester}{\value{refrad}} & I \\
\stepcounter{id}\theid & WR\,146 & high flux density, $\alpha < 0$, resolved NT emitting region & \thedwart,\thewester,\stepcounter{refrad}\therefrad & I \\
\stepcounter{id}\theid & WR\,147 & high flux density, $\alpha < 0$, resolved NT emitting region & \stepcounter{refrad}\therefrad,\thedwart,\thewester,\stepcounter{refrad}\therefrad & I \\
\vspace*{-0.2cm}\\
\hline
\end{tabular}
\end{center}
\tablebib{
\setcounter{refrad}{0}
\stepcounter{refrad}\therefrad. \citet{BAC}; \stepcounter{refrad}\therefrad. \citet{drake}; \stepcounter{refrad}\therefrad. \citet{montes2009}; \stepcounter{refrad}\therefrad. \citet{lei1997}; \stepcounter{refrad}\therefrad. \citet{Chap}; \stepcounter{refrad}\therefrad. \citet{Benaglia2006}; \stepcounter{refrad}\therefrad. \citet{BK2007}; \stepcounter{refrad}\therefrad. \citet{BCK}; \stepcounter{refrad}\therefrad. \citet{BRKP}; \stepcounter{refrad}\therefrad. \citet{BK}; \stepcounter{refrad}\therefrad. \citet{benagliahd93129avlbi}; \stepcounter{refrad}\therefrad. \citet{leietal}; \stepcounter{refrad}\therefrad. \citet{setiaIAU}; \stepcounter{refrad}\therefrad. \citet{Cap}; \stepcounter{refrad}\therefrad. \citet{Mon}; \stepcounter{refrad}\therefrad. \citet{rauw9sgr}; \stepcounter{refrad}\therefrad. \citet{Blo167971}; \stepcounter{refrad}\therefrad. \citet{DeB168112}; \stepcounter{refrad}\therefrad. \citet{Blo168112}; \stepcounter{refrad}\therefrad. \citet{radiocen1}; \stepcounter{refrad}\therefrad. \citet{radiocen12}; \stepcounter{refrad}\therefrad. \citet{ABCT}; \stepcounter{refrad}\therefrad. \citet{williams1992}; \stepcounter{refrad}\therefrad. \citet{scuderi}; \stepcounter{refrad}\therefrad. \citet{DW}; \stepcounter{refrad}\therefrad. \citet{wil140}; \stepcounter{refrad}\therefrad. \citet{wilrad140}; \stepcounter{refrad}\therefrad. \citet{WhBe140}; \stepcounter{refrad}\therefrad. \citet{Doug140}; \stepcounter{refrad}\therefrad. \citet{contr}; \stepcounter{refrad}\therefrad. \citet{kennedycyg5};  \stepcounter{refrad}\therefrad. \citet{ortizcyg5}; \stepcounter{refrad}\therefrad. \citet{vanloocyg9}; \stepcounter{refrad}\therefrad. \citet{vlbi2006}; \stepcounter{refrad}\therefrad. \citet{Blommecyg9}; \stepcounter{refrad}\therefrad. \citet{Blomme8apaper}; \stepcounter{refrad}\therefrad. \citet{setia}; \stepcounter{refrad}\therefrad. \citet{oconnorwr146art};  \stepcounter{refrad}\therefrad. \citet{naturewr147};  \stepcounter{refrad}\therefrad. \citet{williamswr147}}
\end{table*}
\normalsize

\newpage

\setcounter{id}{0}
\newcounter{refhe}
\begin{table*}[ht]
\caption{High energy investigations of PACWBs.\label{highenergy}}
\begin{center}
\begin{tabular}{l l c l l c l l c l}
\hline
\# & usual ID & \multicolumn{2}{c}{Soft X-rays} & & \multicolumn{2}{c}{Hard X-rays} & & \multicolumn{2}{c}{$\gamma$-rays} \\
\cline{3-4}\cline{6-7}\cline{9-10}
 & & Status & Ref. & & Status & Ref. & & Status & Ref. \\
\hline
\vspace*{-0.2cm}\\
\stepcounter{id}\theid & HD\,15558 & D & \stepcounter{refhe}\therefhe\newcounter{chsart}\setcounter{chsart}{\value{refhe}} &  & -- & -- &  & -- & -- \\
\stepcounter{id}\theid & $\delta$Ori\,A & D & \thechsart &  & -- & -- &  & -- & -- \\
\stepcounter{id}\theid & $\sigma$Ori\,AB & D & \thechsart &  & -- & -- &  & -- & -- \\
\stepcounter{id}\theid & 15\,Mon & D & \thechsart &  & -- & -- &  & -- & -- \\
\stepcounter{id}\theid & WR\,8 & D &  &  & -- & -- &  & -- & -- \\
\stepcounter{id}\theid & WR\,11 & D & \stepcounter{refhe}\therefhe\newcounter{phcart}\setcounter{phcart}{\value{refhe}},\stepcounter{refhe}\therefhe,\stepcounter{refhe}\therefhe &  & -- & -- &  & UL & \stepcounter{refhe}\therefhe\newcounter{wernart}\setcounter{wernart}{\value{refhe}} \\
\stepcounter{id}\theid & WR\,14 & D & \thephcart &  & -- & -- &  & -- & -- \\
\stepcounter{id}\theid & CD--47\,4551 & -- & -- &  & -- & -- &  & -- & -- \\
\stepcounter{id}\theid & WR\,21a & D & \stepcounter{refhe}\therefhe,\stepcounter{refhe}\therefhe,\stepcounter{refhe}\therefhe &  & -- & -- &  & -- & -- \\
\stepcounter{id}\theid & HD\,93129A & D & \stepcounter{refhe}\therefhe\newcounter{carart}\setcounter{carart}{\value{refhe}} &  & -- & -- &  & -- & -- \\
\stepcounter{id}\theid & HD\,93250 & D & \thechsart,\stepcounter{refhe}\therefhe,\thecarart &  & -- & -- &  & -- & -- \\
\stepcounter{id}\theid & $\eta$\,Car & D & \stepcounter{refhe}\therefhe,\stepcounter{refhe}\therefhe,\stepcounter{refhe}\therefhe,\stepcounter{refhe}\therefhe,\stepcounter{refhe}\therefhe &  & D & \stepcounter{refhe}\therefhe,\stepcounter{refhe}\therefhe,\stepcounter{refhe}\therefhe &  & D & \stepcounter{refhe}\therefhe,\stepcounter{refhe}\therefhe \\
\stepcounter{id}\theid & WR\,39 & UL & \thephcart &  & -- & -- &  & -- & -- \\
\stepcounter{id}\theid & WR\,48 & D & \thephcart &  & -- & -- &  & -- & -- \\
\stepcounter{id}\theid & HD\,124314 & D & \stepcounter{refhe}\therefhe &  & -- & -- &  & -- & -- \\
\stepcounter{id}\theid & HD\,150136 & D & \stepcounter{refhe}\therefhe &  & -- & -- &  & -- & -- \\
\stepcounter{id}\theid & HD\,151804 & D & \thechsart &  & -- & -- &  & -- & -- \\
\stepcounter{id}\theid & WR\,78 & D & \thephcart &  & -- & -- &  & -- & -- \\
\stepcounter{id}\theid & WR\,79a & D & \stepcounter{refhe}\therefhe &  & -- & -- &  & -- & -- \\
\stepcounter{id}\theid & HD\,152623 & D & \thechsart &  & -- & -- &  & -- & -- \\
\stepcounter{id}\theid & WR\,89 & D & \thephcart &  & -- & -- &  & -- & -- \\
\stepcounter{id}\theid & WR\,90 & -- & -- &  & -- & -- &  & -- & -- \\
\stepcounter{id}\theid & WR\,98 & -- &  &  & -- & -- &  & -- & -- \\
\stepcounter{id}\theid & WR\,98a & -- & -- &  & -- & -- &  & -- & -- \\
\stepcounter{id}\theid & WR\,104 & D & \thephcart &  & -- & -- &  & -- & -- \\
\stepcounter{id}\theid & WR\,105 & D & \thephcart &  & -- & -- &  & -- & -- \\
\stepcounter{id}\theid & 9\,Sgr & D & \thechsart,\stepcounter{refhe}\therefhe &  & -- & -- &  & -- & -- \\
\stepcounter{id}\theid & WR\,112 & UL & \thephcart &  & -- & -- &  & -- & -- \\
\stepcounter{id}\theid & HD\,167971 & D & \thechsart,\stepcounter{refhe}\therefhe &  & -- & -- &  & -- & -- \\
\stepcounter{id}\theid & HD\,168112 & D & \thechsart,\stepcounter{refhe}\therefhe &  & -- & -- &  & -- & -- \\
\stepcounter{id}\theid & CEN\,1a & D & \stepcounter{refhe}\therefhe\newcounter{cenx}\setcounter{cenx}{\value{refhe}} &  & -- & -- &  & -- & -- \\
\stepcounter{id}\theid & CEN\,1b & D & \thecenx &  & -- & -- &  & -- & -- \\
\stepcounter{id}\theid & WR\,125 & D & \thephcart &  & -- & -- &  & UL & \thewernart \\
\stepcounter{id}\theid & HD\,190603 & -- & -- &  & -- & -- &  & -- & -- \\
\stepcounter{id}\theid & WR\,133 & D & \thephcart &  & -- & -- &  & -- & -- \\
\stepcounter{id}\theid & WR\,137 & D & \thephcart &  & -- & -- &  & UL & \thewernart \\
\stepcounter{id}\theid & WR\,140 & D & \thephcart,\stepcounter{refhe}\therefhe,\stepcounter{refhe}\therefhe,\stepcounter{refhe}\therefhe,\stepcounter{refhe}\therefhe\newcounter{sugart}\setcounter{sugart}{\value{refhe}},\stepcounter{refhe}\therefhe &  & D & \thesugart &  & UL & \thewernart \\
\stepcounter{id}\theid & Cyg\,OB2\,\#5 & D & \stepcounter{refhe}\therefhe\newcounter{harart}\setcounter{harart}{\value{refhe}},\stepcounter{refhe}\therefhe\newcounter{kmart}\setcounter{kmart}{\value{refhe}},\stepcounter{refhe}\therefhe\newcounter{rosart}\setcounter{rosart}{\value{refhe}},\stepcounter{refhe}\therefhe\newcounter{chaart}\setcounter{chaart}{\value{refhe}},\stepcounter{refhe}\therefhe,\stepcounter{refhe}\therefhe\newcounter{suzcygart}\setcounter{suzcygart}{\value{refhe}} &  & UL & \stepcounter{refhe}\therefhe\newcounter{intart}\setcounter{intart}{\value{refhe}} &  & -- & -- \\
\stepcounter{id}\theid & Cyg\,OB2\,\#9 & D & \theharart,\thekmart,\therosart,\thechaart,\stepcounter{refhe}\therefhe,\thesuzcygart &  & UL & \theintart &  & -- & -- \\
\stepcounter{id}\theid & Cyg\,OB2\,\#8A & D & \theharart,\thekmart,\therosart,\thechaart,\stepcounter{refhe}\therefhe,\stepcounter{refhe}\therefhe,\thesuzcygart &  & UL & \theintart &  & -- & -- \\
\stepcounter{id}\theid & Cyg\,OB2-335 & -- & -- &  & -- & -- &  & -- & -- \\
\stepcounter{id}\theid & WR\,146 & -- & -- &  & UL & \theintart &  & UL & \thewernart \\
\stepcounter{id}\theid & WR\,147 & D & \thephcart,\stepcounter{refhe}\therefhe,\stepcounter{refhe}\therefhe,\stepcounter{refhe}\therefhe &  & UL & \theintart &  & UL & \thewernart \\
\vspace*{-0.2cm}\\
\hline
\end{tabular}
\end{center}
\footnotesize{Meanings of Status symbols: D stands for detected, UL stands for published uppel limit.
}
\tablebib{
\setcounter{refhe}{0}
\stepcounter{refhe}\therefhe. \citet{CHS1989}; \stepcounter{refhe}\therefhe. \citet{PHC1995}; \stepcounter{refhe}\therefhe. \citet{rauwwr11}; \stepcounter{refhe}\therefhe. \citet{schildgamma2vel}; \stepcounter{refhe}\therefhe. \citet{wernerfermiwr}; \stepcounter{refhe}\therefhe. \citet{caraveowr21a}; \stepcounter{refhe}\therefhe. \citet{mereghettiwr21a}; \stepcounter{refhe}\therefhe. \citet{BRKP}; \stepcounter{refhe}\therefhe. \citet{carinachandra}; \stepcounter{refhe}\therefhe. \citet{rauw93250}; \stepcounter{refhe}\therefhe. \citet{einsteinetacar}; \stepcounter{refhe}\therefhe. \citet{corcoranetacar1998}; \stepcounter{refhe}\therefhe. \citet{sewardetacar}; \stepcounter{refhe}\therefhe. \citet{ishibashietacar}; \stepcounter{refhe}\therefhe. \citet{corcoranetacar2010}; \stepcounter{refhe}\therefhe. \citet{viottietacar}; \stepcounter{refhe}\therefhe. \citet{leyderetacar}; \stepcounter{refhe}\therefhe. \citet{sekiguchietacar}; \stepcounter{refhe}\therefhe. \citet{tavanietacar}; \stepcounter{refhe}\therefhe. \citet{farnieretacar}; \stepcounter{refhe}\therefhe. \citet{vogesrosat}; \stepcounter{refhe}\therefhe. \citet{Xray150136}; \stepcounter{refhe}\therefhe. \citet{Xraywr79a}; \stepcounter{refhe}\therefhe. \citet{rauw9sgr}; \stepcounter{refhe}\therefhe. \citet{DeB167971}; \stepcounter{refhe}\therefhe. \citet{DeB168112}; \stepcounter{refhe}\therefhe. \citet{broosm17}; \stepcounter{refhe}\therefhe. \citet{koyama140}; \stepcounter{refhe}\therefhe. \citet{ZS140}; \stepcounter{refhe}\therefhe. \citet{pollock140}; \stepcounter{refhe}\therefhe. \citet{sugawara140}; \stepcounter{refhe}\therefhe. \citet{debeckerliacwr140}; \stepcounter{refhe}\therefhe. \citet{harnden}; \stepcounter{refhe}\therefhe. \citet{KM}; \stepcounter{refhe}\therefhe. \citet{cygrosat}; \stepcounter{refhe}\therefhe. \citet{cygchandra}; \stepcounter{refhe}\therefhe. \citet{Lindercyg5}; \stepcounter{refhe}\therefhe. \citet{cygob2suzaku}; \stepcounter{refhe}\therefhe. \citet{cygint}; \stepcounter{refhe}\therefhe. \citet{nazecyg9}; \stepcounter{refhe}\therefhe. \citet{DeBcyg8a}; \stepcounter{refhe}\therefhe. \citet{Blomme8apaper}; \stepcounter{refhe}\therefhe. \citet{wr147pittard}; \stepcounter{refhe}\therefhe. \citet{skinner-wr147}; \stepcounter{refhe}\therefhe. \citet{wr147zhekov}.
}
\end{table*}
\normalsize

\newpage

\setcounter{id}{0}
\newcounter{refdisc}
\newcounter{refdiscb}
\begin{table*}[ht]
\caption{Adopted parameters and calculated wind kinetic power and radio luminosities.\label{param}}
\begin{center}
\begin{tabular}{l l l l l l l l l l l}
\hline
\# & usual ID & Sp. type & ${\mathrm{\dot M}}$ & V$_\infty$ & P$_\mathrm{kin}$ & S$_\nu$ & $\lambda$ & $\alpha$ & Ref. & L$_\mathrm{rad}$\\
  &  &  & (M$_\odot$\,yr$^{-1}$) & (km\,s$^{-1}$) & (erg\,s$^{-1}$) & (mJy) & (cm) & & & (erg\,s$^{-1}$) \\
\hline
\vspace*{-0.2cm}\\
\stepcounter{id}\theid & HD\,15558 & O5.5III(f) & 1.4\,$\times$\,10$^{-6}$ & 2900 &  3.7\,$\times$\,10$^{36}$& 0.5 & 6.0 & (0.0) & \stepcounter{refdiscb}\therefdiscb\newcounter{fluxbac}\setcounter{fluxbac}{\value{refdiscb}} & (9.4\,$\times$\,10$^{28}$) \\
\stepcounter{id}\theid & $\delta$Ori\,A & O9.5II & 2.3\,$\times$\,10$^{-7}$ & 2800 & 5.7\,$\times$\,10$^{35}$& 0.37 & 6.0 & (0.0) & \thefluxbac & (2.2\,$\times$\,10$^{27}$) \\
\stepcounter{id}\theid & $\sigma$Ori\,AB & O9.5V & 1.6\,$\times$\,10$^{-8}$ & 2300 & 2.7\,$\times$\,10$^{34}$& 1.78 & 6.0 & --0.4 & \stepcounter{refdiscb}\therefdiscb\newcounter{fluxdr}\setcounter{fluxdr}{\value{refdiscb}} & 9.4\,$\times$\,10$^{27}$ \\
\stepcounter{id}\theid & 15\,Mon & O7V(f) & 4.6\,$\times$\,10$^{-8}$ & 2500 & 9.1\,$\times$\,10$^{34}$& 0.4 & 6.0 & (0.0) & \thefluxdr & (1.3\,$\times$\,10$^{28}$) \\
\stepcounter{id}\theid & WR\,8 & WN7 & 8.0\,$\times$\,10$^{-5}$ & 1200 & 3.6\,$\times$\,10$^{37}$& 0.22 & 6.0 & 0.17 & \thefluxdr & 1.1\,$\times$\,10$^{29}$ \\
\stepcounter{id}\theid & WR\,11 & WC8 & 1.6\,$\times$\,10$^{-5}$ & 1800 & 1.6\,$\times$\,10$^{37}$ & 26.5 & 6.0 & 0.3 & \stepcounter{refdiscb}\therefdiscb\newcounter{fluxlei}\setcounter{fluxlei}{\value{refdiscb}} & 1.5\,$\times$\,10$^{29}$ \\
\stepcounter{id}\theid & WR\,14 & WC7 & 1.6\,$\times$\,10$^{-5}$ & 2000 & 2.0\,$\times$\,10$^{37}$& 0.46 & 6.0 & --0.82 & \thefluxlei & 4.7\,$\times$\,10$^{28}$ \\
\stepcounter{id}\theid & CD--47\,4551 & O5If & 2.7\,$\times$\,10$^{-6}$ & 3000 & 7.7\,$\times$\,10$^{36}$ & 2.98 & 6.0 & --0.75 & \stepcounter{refdiscb}\therefdiscb & 1.3\,$\times$\,10$^{29}$ \\
\stepcounter{id}\theid & WR\,21a & O3If$^*$/WN6ha & 1.3\,$\times$\,10$^{-5}$ & 2200 & 2.0\,$\times$\,10$^{37}$ & 0.25 & 6.0 & (0.0) & \stepcounter{refdiscb}\therefdiscb & (8.0\,$\times$\,10$^{28}$) \\
\stepcounter{id}\theid & HD\,93129A & O2If$^*$ & 5.0\,$\times$\,10$^{-6}$ & 3300 & 1.7\,$\times$\,10$^{37}$ & 4.1 & 6.0 & --1.04 & \stepcounter{refdiscb}\therefdiscb & 4.5\,$\times$\,10$^{29}$ \\
\stepcounter{id}\theid & HD\,93250 & O4III & 2.9\,$\times$\,10$^{-6}$ & 2900 & 7.7\,$\times$\,10$^{36}$ & 1.36 & 3.6 & (0.0) & \stepcounter{refdiscb}\therefdiscb & (2.7\,$\times$\,10$^{29}$) \\
\stepcounter{id}\theid & $\eta$\,Car & LBV? & 2.5\,$\times$\,10$^{-4}$ & 600 & 2.8\,$\times$\,10$^{37}$ & -- & -- & -- & -- & -- \\
\stepcounter{id}\theid & WR\,39 & WC7 & 1.6\,$\times$\,10$^{-5}$ & 2000 & 2.0\,$\times$\,10$^{37}$ & 0.87 & 6.0 & --0.31 & \stepcounter{refdiscb}\therefdiscb\newcounter{fluxcha}\setcounter{fluxcha}{\value{refdiscb}} & 8.1\,$\times$\,10$^{29}$ \\
\stepcounter{id}\theid & WR\,48 & WC5 & 2.4\,$\times$\,10$^{-5}$ & 2800 & 5.9\,$\times$\,10$^{37}$ & 2.06 & 6.0 & --0.33 & \thefluxcha & 3.3\,$\times$\,10$^{29}$ \\
\stepcounter{id}\theid & HD\,124314 & O6V(n)((f)) & 4.1\,$\times$\,10$^{-7}$ & 2600 & 8.7\,$\times$\,10$^{35}$ & 4.14 & 6.0 & --0.64 & \stepcounter{refdiscb}\therefdiscb\newcounter{fluxben}\setcounter{fluxben}{\value{refdiscb}} & 1.1\,$\times$\,10$^{29}$ \\
\stepcounter{id}\theid & HD\,150136 & O3-3.5V((f$^+$)) & 2.3\,$\times$\,10$^{-6}$ & 2700 & 5.3\,$\times$\,10$^{36}$ & 5.57 & 6.0 & --1.29 & \thefluxben & 3.2\,$\times$\,10$^{29}$ \\
\stepcounter{id}\theid & HD\,151804 & O8Iaf & 8.7\,$\times$\,10$^{-7}$ & 2400 & 1.6\,$\times$\,10$^{36}$ & 0.8 & 6.0 & --0.18 & \stepcounter{refdiscb}\therefdiscb\newcounter{fluxset}\setcounter{fluxset}{\value{refdiscb}} & 6.9\,$\times$\,10$^{28}$ \\
\stepcounter{id}\theid & WR\,78 & WN7h & 6.3\,$\times$\,10$^{-5}$ & 1200 & 2.9\,$\times$\,10$^{37}$ & 1.1 & 6.0 & 0.31 & \thefluxset & 1.3\,$\times$\,10$^{29}$ \\
\stepcounter{id}\theid & WR\,79a & WN9ha & 2.5\,$\times$\,10$^{-5}$ & 900 & 6.4\,$\times$\,10$^{36}$ & 0.8 & 6.0 & 0.31 & \thefluxset & 9.8\,$\times$\,10$^{28}$ \\
\stepcounter{id}\theid & HD\,152623 & O7V((f)) & 4.6\,$\times$\,10$^{-8}$ & 2500 & 9.1\,$\times$\,10$^{34}$ & 0.5 & 6.0 & --0.94 & \thefluxset & 3.4\,$\times$\,10$^{28}$ \\
\stepcounter{id}\theid & WR\,89 & WN8h & 8.0\,$\times$\,10$^{-5}$ & 1600 & 6.5\,$\times$\,10$^{37}$ & 2.0 & 3.6 & (0.0) & \stepcounter{refdiscb}\therefdiscb & (7.7\,$\times$\,10$^{29}$) \\
\stepcounter{id}\theid & WR\,90 & WC7 & 1.6\,$\times$\,10$^{-5}$ & 2000 & 2.0\,$\times$\,10$^{37}$ & 1.09 & 6.0 & --0.01 & \thefluxcha & 1.0\,$\times$\,10$^{29}$ \\
\stepcounter{id}\theid & WR\,98 & WN7/WC & 6.3\,$\times$\,10$^{-5}$ & 2100 & 8.8\,$\times$\,10$^{37}$& 0.94 & 6.0 & 0.26 & \thefluxdr & 1.5\,$\times$\,10$^{29}$ \\
\stepcounter{id}\theid & WR\,98a & WC9 & 1.6\,$\times$\,10$^{-5}$ & 1400 & 9.9\,$\times$\,10$^{36}$ & 0.62 & 3.6 & 0.05 & \stepcounter{refdiscb}\therefdiscb\newcounter{fluxmon}\setcounter{fluxmon}{\value{refdiscb}} & 8.1\,$\times$\,10$^{28}$ \\
\stepcounter{id}\theid & WR\,104  & WC9 & 1.6\,$\times$\,10$^{-5}$ & 1400 & 9.9\,$\times$\,10$^{36}$ & 0.87 & 3.6 & 0.28 & \thefluxmon & 8.9\,$\times$\,10$^{28}$ \\
\stepcounter{id}\theid & WR\,105 & WN9h & 6.3\,$\times$\,10$^{-5}$ & 700 & 9.7\,$\times$\,10$^{36}$ & 4.39 & 6.0 & --0.3 & \thefluxlei & 3.1\,$\times$\,10$^{29}$ \\
\stepcounter{id}\theid & 9\,Sgr & O3.5V((f$^+$)) & 1.9\,$\times$\,10$^{-6}$ & 2700 & 4.4\,$\times$\,10$^{36}$& 2.8 & 6.0 & --1.10 & \stepcounter{refdiscb}\therefdiscb & 2.0\,$\times$\,10$^{29}$ \\
\stepcounter{id}\theid & WR\,112 & WC9 & 1.6\,$\times$\,10$^{-5}$ & 1400 & 9.9\,$\times$\,10$^{36}$ & 4.07 & 3.6 & 0.13 & \thefluxmon & 2.6\,$\times$\,10$^{30}$ \\
\stepcounter{id}\theid & HD\,167971 & O8I & 8.7\,$\times$\,10$^{-7}$ & 2400 & 2.0\,$\times$\,10$^{36}$ & 17.1 & 6.0 & --0.48 & \stepcounter{refdiscb}\therefdiscb & 1.3\,$\times$\,10$^{30}$ \\
\stepcounter{id}\theid & HD\,168112 & O5.5III(f$^+$) & 1.4\,$\times$\,10$^{-6}$ & 2900 & 3.7\,$\times$\,10$^{36}$ & 5.64 & 6.0 & --0.78 & \stepcounter{refdiscb}\therefdiscb & 4.1\,$\times$\,10$^{29}$ \\
\stepcounter{id}\theid & CEN\,1a & O4V &  1.5\,$\times$\,10$^{-6}$ & 2600 & 3.2\,$\times$\,10$^{36}$ & 2.69 & 6.0 & --0.08 & \stepcounter{refdiscb}\therefdiscb\newcounter{cenref}\setcounter{cenref}{\value{refdiscb}} & 3.9\,$\times$\,10$^{29}$ \\
\stepcounter{id}\theid & CEN\,1b & O4V &  1.5\,$\times$\,10$^{-6}$ & 2600 & 3.2\,$\times$\,10$^{36}$ & 1.15 & 6.0 & --0.68 & \thecenref & 1.3\,$\times$\,10$^{29}$ \\
\stepcounter{id}\theid & WR\,125 & WC7 & 1.6\,$\times$\,10$^{-5}$ & 2000 & 2.0\,$\times$\,10$^{37}$ & 1.5 & 6.0 & --0.57 & \stepcounter{refdiscb}\therefdiscb & 1.5\,$\times$\,10$^{29}$ \\
\stepcounter{id}\theid & HD\,190603 & B1.5Ia & 2.0\,$\times$\,10$^{-6}$ & 500 & 1.6\,$\times$\,10$^{35}$ & 0.6 & 6.0 & (0.0) & \stepcounter{refdiscb}\therefdiscb & (4.8\,$\times$\,10$^{28}$) \\
\stepcounter{id}\theid & WR\,133 & WN5 & 4.0\,$\times$\,10$^{-5}$ & 1400 & 2.5\,$\times$\,10$^{37}$& 0.38 & 6.0 & --0.65 & \thefluxdr & 4.4\,$\times$\,10$^{28}$ \\
\stepcounter{id}\theid & WR\,137 & WC7 & 1.6\,$\times$\,10$^{-5}$ & 2000 & 2.0\,$\times$\,10$^{37}$ & -- & -- & 0.0 & \stepcounter{refdiscb}\therefdiscb & -- \\
\stepcounter{id}\theid & WR\,140 & WC7 & 2.0\,$\times$\,10$^{-5}$ & 3100 & 6.1\,$\times$\,10$^{37}$ & 26.8 & 6.0 & --0.27 & \stepcounter{refdiscb}\therefdiscb & 2.6\,$\times$\,10$^{30}$ \\
\stepcounter{id}\theid & Cyg\,OB2\,\#5 & Ofpe/WN9 & 1.3\,$\times$\,10$^{-5}$ & 2000 & 1.6\,$\times$\,10$^{37}$ & 2.98 & 6.0 & --0.06 & \stepcounter{refdiscb}\therefdiscb & 2.9\,$\times$\,10$^{29}$ \\
\stepcounter{id}\theid & Cyg\,OB2\,\#9 & O5I & 2.7\,$\times$\,10$^{-6}$ & 3000 & 7.7\,$\times$\,10$^{36}$ & 6.2 & 6.0 & --0.38 & \stepcounter{refdiscb}\therefdiscb & 4.9\,$\times$\,10$^{29}$ \\
\stepcounter{id}\theid & Cyg\,OB2\,\#8A & O6If & 1.8\,$\times$\,10$^{-6}$ & 3000 & 5.1\,$\times$\,10$^{36}$ & 1.33 & 6.0 & --0.22 & \stepcounter{refdiscb}\therefdiscb & 1.2\,$\times$\,10$^{29}$ \\
\stepcounter{id}\theid & Cyg\,OB2-335 & O7V & 4.6\,$\times$\,10$^{-8}$ & 2500 & 9.1\,$\times$\,10$^{34}$ & 1.3 & 6.0 & --0.8 & \stepcounter{refdiscb}\therefdiscb & 9.6\,$\times$\,10$^{28}$ \\
\stepcounter{id}\theid & WR\,146 & WC6 & 2.5\,$\times$\,10$^{-5}$ & 2300 & 4.2\,$\times$\,10$^{37}$ & 33 & 6.0 & --0.6 & \stepcounter{refdiscb}\therefdiscb & 1.2\,$\times$\,10$^{30}$ \\
\stepcounter{id}\theid & WR\,147 & WN8 & 5.0\,$\times$\,10$^{-5}$ & 1000 & 1.6\,$\times$\,10$^{37}$ & 38 & 6.0 & 0.05 & \stepcounter{refdiscb}\therefdiscb & 6.0\,$\times$\,10$^{29}$ \\
\vspace*{-0.2cm}\\
\hline
\end{tabular}
\end{center}
\setcounter{refdisc}{0}
\tablebib{
\setcounter{refdiscb}{0}
\stepcounter{refdiscb}\therefdiscb. \citet{BAC}; \stepcounter{refdiscb}\therefdiscb. \citet{drake}; \stepcounter{refdiscb}\therefdiscb. \citet{lei1997}; \stepcounter{refdiscb}\therefdiscb. \citet{BCK}; \stepcounter{refdiscb}\therefdiscb. \citet{BRKP}; \stepcounter{refdiscb}\therefdiscb. \citet{BK}; \stepcounter{refdiscb}\therefdiscb. \citet{leietal}; \stepcounter{refdiscb}\therefdiscb. \citet{Chap}; \stepcounter{refdiscb}\therefdiscb. \citet{Benaglia2006}; \stepcounter{refdiscb}\therefdiscb. \citet{setiaIAU}; \stepcounter{refdiscb}\therefdiscb. \citet{Cap}; \stepcounter{refdiscb}\therefdiscb. \citet{Mon}; \stepcounter{refdiscb}\therefdiscb. \citet{rauw9sgr}; \stepcounter{refdiscb}\therefdiscb. \citet{Blo167971}; \stepcounter{refdiscb}\therefdiscb. \citet{Blo168112}; \stepcounter{refdiscb}\therefdiscb. \citet{radiocen12}; \stepcounter{refdiscb}\therefdiscb. \citet{williams1992}; \stepcounter{refdiscb}\therefdiscb. \citet{scuderi}; \stepcounter{refdiscb}\therefdiscb. \citet{DW}; \stepcounter{refdiscb}\therefdiscb. \citet{WhBe140}; \stepcounter{refdiscb}\therefdiscb. \citet{kennedycyg5}; \stepcounter{refdiscb}\therefdiscb. \citet{vanloocyg9}; \stepcounter{refdiscb}\therefdiscb. \citet{Blomme8apaper}; \stepcounter{refdiscb}\therefdiscb. \citet{setia}; \stepcounter{refdiscb}\therefdiscb. \citet{wr146setia}; \stepcounter{refdiscb}\therefdiscb. \citet{naturewr147}.}
\end{table*}
\normalsize

\newpage

\section{Notes on List\,A objects}\label{individualA}

\paragraph{HD\,15558}
The primary object in system presents an unexpectedly high minimum mass, suggesting that the primary object might be an unrevealed close binary \citep{ic1805_2}. The non-thermal radio emission should come from the interaction between the objects orbiting with a period of about 442\,d. HD\,15558 never benefited from radio monitoring or a dedicated X-ray observation.

\paragraph{$\delta$\,Ori\,A}
In this triple system, the synchrotron radio emission is not expected to arise from the wind-wind interaction in the close O + B system but rather from the interaction between the close system and the third and distant star orbiting with a very long period.

\paragraph{$\sigma$\,Ori\,AB}
This multiple system contains at least 3 OB stars, allowing multiple possibilities for the production of synchrotron radio emission. A clarification of the complex multiplicity of this object is strongly needed to discuss the physics of its particle acceleration.

\paragraph{15\,Mon}
The long period of this O + O system (about 25\,yr, \citealt{gies15mon}) is compatible with the escape of synchrotron radio photons, which agrees with the results of radio observations. 

\paragraph{WR\,8}
This WN/C system has been reported as a spectroscopic binary with a period of 38.4\,d by \citet{niemela1991binwr}, but \citet{wr8bin1998} reported on a period of about 115\,d on the basis of Hipparcos photometric data. 

\paragraph{WR\,11}
This WC + O binary is one of a few CWBs that has been observed using long baseline near-infrared interferometry \citep{millourwr11}. However, the short separation between the two components in the system did not allow the authors to resolve them. The 'not so low' radio spectral index ($\alpha \sim 0.3$) is most probably explained by the short period, leading a significant fraction of the synchrotron radio emission component to be absorbed by the stellar winds (mainly, that of the WC star). This is the nearest system in the catalogue.

\paragraph{WR\,14}
A significant variability has been reported in the radio domain for this object, which is most likely attributable to binarity, even though its multiplicity status still needs clarification. A variability in the radio polarization has also been reported, suggesting significant deviation from spherical symmetry, lending further support to the binary candidate status \citep{wr14pola}.

\paragraph{CD--47\,4551}
This is undoubtedly a non-thermal radio emitter with a well-established high flux density and a negative spectral index. This system never benefited from a dedicated multiplicity study. It is recommended to perform such a study.

\paragraph{WR\,21a}
A close WN + O eccentric binary that would deserve radio monitoring to adequately quantify its non-thermal contribution. Its large eccentricity (e = 0.64) suggests significant variations in the conditions ruling the production (and the escape) of synchrotron photons as a function of the orbital phase.

\paragraph{HD\,93129A}
This object benefited from dedicated radio observations, confirming to a large extent its non-thermal radio emitter status. Preliminary VLBI results show a resolved elongated radio emission region most likely coincident with the wind-wind interaction region \citep{benagliahd93129avlbi}. This system is the earliest-type O + O system listed in the catalogue.

\paragraph{HD\,93250}
This long (undetermined) period binary has been resolved by the VLTI, providing first evidence that this is not a single star \citep{hd93250vlti}. The period is undetermined but should be of at least a few 100\,d. On the other hand, additional radio observations are needed to ascertain its non-thermal radio emitter status.

\paragraph{$\eta$\,Car}
Indirect evidence of binarity have been provided by \citet{etacarbin}, but final confirmation is still lacking. This object is the only PACWB that has been detected as a hard X-ray and $\gamma$-ray emitter, whilst no evidence of synchrotron radio emission has been reported. The detection of $\gamma$-rays provided compelling evidence for the first time that PACWBs may be the scene of hadronic processes \citep[see e.g.][]{farnieretacar}. The lack of non-thermal emission detection is attributed to the strong free-free absorption by the dense circumstellar material. Even though a significant variability has been reported in the radio domain, this is not attributed to a non-thermal contribution \citep{etacarradio}.

\paragraph{WR\,39}
The non-thermal emission component from this object presents a significant offset with respect to the expected position of the WC star, strongly suggesting its association with a wind-wind interaction ragion in a binary system \citep{Chap}. However, the binary status of WR\,39 still needs confirmation.

\paragraph{WR\,48}
In this triple system, the bulk of the non-thermal radio emission is a priori expected to arise from the wind-wind interaction involving the third more distant component, orbiting with an undetermined period. The 19\,d period of the harder WC + O binary is too short to allow a significant synchrotron emission to escape from the wind-wind interaction in between, considering the thickness of the WC stellar wind. 

\paragraph{HD\,124314}
Hints for radial velocity shifts attributable to the presence of a companion have been reported by \citet{feasthd124314}, and the presence of a visual companion at about 2.7\,arcsecond has been reported by \citet{WDScata}. However, it is not clear whether these two results are directly related. A multiplicity investigation is needed to clarify its nature and derive valuable orbital elements.

\paragraph{HD\,150136}
In this triple system, the non-thermal radio emission is most probably emerging from the wind-wind interaction region due to the presence of the third star \citep{mahyhd150136,sana150136}. This is the first system in the catalogue with a 3D orbit determination on the basis of spectroscopic and interferometric results.

\paragraph{HD\,151804}
The multiplicity of this O-star has never been investigated.

\paragraph{WR\,78}
The multiplicity for this WN system has never been investigated. This object has been detected in soft X-rays.

\paragraph{WR\,79a}
The presence of a companion has been reported by \citet{masonspeckle} through speckle observations. The companion should be located at a few arcseconds, pointing to a very long period orbit. It has been proposed to be a transition object between O-type and WN-type by \citet{crowther1997}.

\paragraph{HD\,152623}
The presence of a companion has been reported by \citet{masonspeckle} through speckle observations. The companion should be located at a fraction of arcsecond, pointing to a long period orbit.

\paragraph{WR\,89}
This is a variable radio source that has been detected in soft X-rays. It is classified as a WN + OB binary by \citet{vdhcata}, but the nature of the companion and the orbital period are not established yet.

\paragraph{WR\,90}
The multiplicity for this WC system has never been investigated so far, and it has not been reported as a soft X-ray source.

\paragraph{WR\,98}
The study by \citet{niemela1991binwr} revealed the binary nature of WR\,98 with a period of 48.7\,d.

\paragraph{WR\,98a}
A well-established dust maker producing a pinwheel nebula. The spectral classification of the companion is not yet fully determined. WR98a has so far never been detected in the high energy domain.

\paragraph{WR\,104}
This system is the emblematic pinwheel nebula, which is almost viewed pole-on, with an orbital period of about 220\,d.

\paragraph{WR\,105} 
This WN object never benefited from a dedicated multiplicity investigation, but its brightness in X-rays suggests an emission level above that expected from a single star \citep{Chap}.

\paragraph{9\,Sgr}
This is a long period, eccentric massive binary, whose binary nature has been confirmed in both spectroscopy \citep{rauw9sgr,rauw9sgr2} and interferometry (Sana\,2011, private communication).

\paragraph{WR\,112}
Indirect evidence of binarity has been provided through modelling of dust production based on near-infrared observations \citep{marchenkowr112}. However, direct detection of the companion is still lacking.

\paragraph{HD\,167971}
This system harbours a close binary and a third star, whose gravitational link with the close binary has recently been revealed by interferometric observation with the VLTI \citep{vlti167971}. Radio light curves suggest a period of about 20 years \citep{Blo167971}, although further near-infrared interferometric observations are needed to determine orbital parameters.

\paragraph{HD\,168112}
The binarity of HD\,168112 has never been confirmed by spectroscopic studies \citep{multjenam}, even though convincing indirect evidence is provided by X-ray \citep{DeB168112} and radio \citep{Blo168112} observations. These results motivate to consider interferometric observations aiming at detecting the still unrevealed companion.

\paragraph{CEN\,1a}
The system CEN\,1 (Kleinmann's Star, \citealt{ChiniM17}) has been identified as a multiple system, made of two massive binaries separated by about 1.8\,arcsec \citep{hoffmeister2008}. Components a and b of this trapezium-like system appear separately in the catalogue, because both binaries are distinct non-thermal radio emitters \citep{radiocen12}, strongly suggesting that the two wind-wind interaction regions act separately as particle accelerators. Component a is variable in the radio domain, suggesting an eccentric orbit.

\paragraph{CEN\,1b}
See CEN\,1a. Component b has not been reported to be a variable radio source, but its radio spectral index definitely points to the existence of a non-thermal emission component \citep{radiocen12}.

\paragraph{WR\,125}
This object is a well-established colliding-wind binary with a long period (several years). However, a clarification of its orbital elements is still needed.

\paragraph{HD\,190603}
A multiplicity investigation is still lacking for this object, that is the lastest type star listed in the catalogue. This star presents properties reminiscent of LBV-like objects \citep{clarkBHG}.

\paragraph{WR\,133}
\citet{montes2009} reported on a composite thermal/non-thermal radio spectrum for WR\,133. However, it should be noted that the short orbital period ($\sim$\,112\,d, \citealt{UH1994wr133}) suggests the wind-wind interaction region would be located deep in the radio-sphere of the strong WR stellar wind, therefore significantly reducing the amount of detectable synchrotron radiation. 

\paragraph{WR\,137}
A long-period ($\sim$\,13\,yr) dust-producing WC + O binary with well-defined orbital elements \citep{wr137orbit}, whose detection in the soft X-ray domain is probably attributable to colliding-winds. 

\paragraph{WR\,140} 
This WC + O system is undoubtedly the colliding-wind binary whose parameters have been established with the highest accuracy. The previous periastron passage (January 2009) triggered a special effort to improve our knowledge of this 7.9\,yr period system \citep{WilliamsLIAC}. Among other results, the study by \citet{fahed140mnras} significantly improved the orbital parameters. WR\,140 is unique in the sense that it is the only PACWB that presents detectable non-thermal emission both in the radio and in the hard X-ray domains. In addition, its non-thermal radio emission has been resolved at several orbital phases, allowing a significant improvement of the modelling of this system both in the radio and the high energy domains \citep{Doug140,PD140,PD140art,Pit2}. 

\paragraph{Cyg\,OB2\,\#5}
Until a few years ago, this system was known to be a triple system, made of a close early-type binary with a much more distant B-type third component. The presence of a fourth star in this multiple system is strongly suggested by a well-defined radio variability (of non-thermal origin) on a time scale of about 6.7 years \citep{kennedycyg5}. Some additional convincing information comes from X-ray observations revealing a hard spectrum in the soft X-ray domain \citep{Lindercyg5}. This last property is most likely attributable to a wind-wind interaction in a binary with a period of a few years, rather than in a binary with a short period of 6.6\,d or a much longer one of more than 9000\,yr. Finally, the recent detection of a radio emitting region at about 12\,mas away from the close binary \citep{ortizcyg5} lends further support to the presence of a colliding-wind region associated with the 6.7\,yr orbit.  

\paragraph{Cyg\,OB2\,\#9}
Recent studies revealed the binary nature of this object \citep{vanloocyg9,nazecyg9}, confirming that the non-thermal radio emission should arise from the wind-wind interaction between the two O stars in this eccentric long period ($\sim$\,2.35\,yr) binary.

\paragraph{Cyg\,OB2\,\#8A}
This massive binary with a period of 21.908\,d \citep{Let8a} is one of the colliding-wind binaries with the best-constrained stellar, wind, and orbital parameters so far. It has also been intensively investigated in X-rays with a well-studied phase-locked variability \citep{DeBcyg8a,Blomme8apaper} and upper limits on the hard X-ray emission derived from {\it INTEGRAL} observations \citep{cygint}. The detailed study by \citet{Blomme8apaper} reported on a phase-dependent radio light curve and presented some modelling of the non-thermal radio emission from Cyg\,OB2\,\#8A. This reveals a significant impact of free-free absorption on the radio light curve and suggests a revision of the stellar wind parameters of the two stars in the system. 

\paragraph{Cyg\,OB2-335}
Even though this object is known as a binary system \citep{kiminkicygob2}, its short period of only a few days is puzzling. The short period indeed translates into a short stellar separation, which is likely uncompatible with a significant emerging synchrotron radio flux. The detection of a significant non-thermal radio contribution from this system \citep{setia} suggests a third undetected star that may orbit on a wider orbit (with a longer period). The idea that Cyg\,OB2-335 would consist of a hierarchized triple system would make it similar to cases such as of HD\,167971 and HD\,150136. In contrast with these two systems, Cyg\,OB2-335 has not yet been detected in soft X-rays.

\paragraph{WR\,146}
Long baseline radio observations allowed \citet{oconnorwr146art} to resolve the non-thermal emitting region. The period of the system is expected to be long (i.e. many years) and is undetermined. 

\paragraph{WR\,147}
This is the only PACWB whose wind-wind interaction region has been resolved in X-rays \citep{wr147pittard}, and the synchrotron emission region has also been resolved in the radio domain \citep{williamswr147}. Its very long period is undetermined.

\section{Notes on List\,B objects}\label{individualB}

\paragraph{CC\,Cas}
This object is a radio variable massive binary \citep{CCCas}, but the short period ($\sim$\,3\,d) does not allow us to reject an eclipse-like variation in the free-free absorption without invoking non-thermal processes.

\paragraph{$\xi$\,Per}
Non-thermal emission of $\xi$\,Per is reported by \citet{Schnerr2007} but since it is a probable runaway star, it might be related to the bow-shock of the stellar wind in interaction with the interstellar medium. 

\paragraph{$\alpha$\,Cam}
Variable flux density at 5\,GHz of $\alpha$\,Cam is quoted by \citet{radio250ghz}, but it is classified as constant by \citet{Schnerr2007}. In addition, as $\xi$\,Per, it is a runaway star candidate.

\paragraph{$\delta$\,Ori\,C}
This object is classified by \citet{drake1987} as a non-thermal emitte, but it is not clear at all whether the acceleration process is related to DSA mechanism in CWBs. This object is a magnetic Bp star.

\paragraph{HD\,37017}
Same comment as for $\delta$\,Ori\,C.

\paragraph{$\theta^1$\,Ori\,A}
This object is a known non-thermal emitter whose nature is not fully clarified. It presents flaring events in probable relation with the presence of a T\,Tauri star component \citep{thetaoriNT}.

\paragraph{$\sigma$\,Ori\,E}
Same comment as for $\delta$\,Ori\,C.

\paragraph{WR\,22}
The radio spectral index has been measrued to lie between 0.0 and 0.6 with a large error bar compatible with thermal or non-thermal emission \citep{leietal}. 

\paragraph{WR\,79}
\citet{setiaIAU} reported on non-thermal radio emission, but it might be attributable to another point source considering the crowded region where this object is located. Notably, \citet{zhekov2012} reported on the existence of two close X-ray emitting sources at a few arcsec, possibly related to pre-main sequence objects.

\paragraph{WR\,86}
The radio spectral index has been measured to lie between 0.0 and 0.6 with a large error bar compatible with thermal or non-thermal emission \citep{DW}. 

\paragraph{W43\,\#1}
A non-thermal radio emission has been reported for this object by \citet{w43nt}, but it is not clear whether it comes from the WR binary itself or a combination of phenomena in the W43 cluster.

\paragraph{Cyg\,OB2\,\#11}
This object is reported as a non-thermal radio emitter by \citet{benagliareview}, but no clear evidence is found in the paper by \citet{BAC}.

\paragraph{WR\,156}
\citet{montes2009} classified WR\,156 as a composite source (thermal + non-thermal) on the basis of a radio spectral index lower than 0.6. However, the spectral index value (0.46\,$\pm$\,0.05) is too high with respect to the criterion ($\alpha < 0.3$) adopted in this study. One can therefore not reject a thermal contribution from the colliding-winds to be responsible for this value.

\end{document}